\documentclass[fleqn,usenatbib]{mnras}

\usepackage{newtxtext,newtxmath}

\usepackage[T1]{fontenc}
\usepackage{ae,aecompl}

\usepackage[usenames,dvipsnames]{xcolor}

\usepackage{graphicx}	
\usepackage{amsmath}	
\usepackage{amssymb}	

\title[Star-formation versus arm number]{Galaxy Zoo: star-formation versus spiral arm number}

\author[Hart et al.]{Ross~E.~Hart,$^1$\thanks{E-mail: ross.hart@nottingham.ac.uk} Steven~P.~Bamford,$^1$ Kevin R.V. Casteels,$^2$ Sandor. J. Kruk,$^3$ \newauthor Chris J. Lintott,$^3$ Karen L. Masters$^4$ 
\\
$^{1}$School of Physics \& Astronomy, The University of Nottingham, University Park, Nottingham NG7 2RD, UK\\
$^{2}$Department of Physics and Astronomy, University of Victoria, Victoria, BC V8P 1A1, Canada \\
$^{3}$Oxford Astrophysics, Department of Physics, University of Oxford, Denys Wilkinson Building, Keble Road, Oxford OX1 3RH, UK\\
$^{4}$Institute for Cosmology and Gravitation, University of Portsmouth, Dennis Sciama Building, Portsmouth PO1 3FX, UK}

\date{Accepted XXX. Received YYY; in original form ZZZ}

\pubyear{2017}

\begin{document}
\label{firstpage}
\pagerange{\pageref{firstpage}--\pageref{lastpage}}
\maketitle

\begin{abstract}
Spiral arms are common features in low-redshift disc galaxies, and are prominent sites of star-formation and dust obscuration. However, spiral structure can take many forms: from galaxies displaying two strong `grand design' arms, to those with many `flocculent' arms. We investigate how these different arm types are related to a galaxy's star-formation and gas properties by making use of visual spiral arm number measurements from Galaxy Zoo 2. We combine UV and mid-IR photometry from GALEX and WISE to measure the rates and relative fractions of obscured and unobscured star formation in a sample of low-redshift SDSS spirals. Total star formation rate has little dependence on spiral arm multiplicity, but two-armed spirals convert their gas to stars more efficiently. We find significant differences in the fraction of obscured star-formation: an additional $\sim 10$ per cent of star-formation in two-armed galaxies is identified via mid-IR dust emission, compared to that in many-armed galaxies. The latter are also significantly offset below the IRX-$\beta$ relation for low-redshift star-forming galaxies. We present several explanations for these differences versus arm number: variations in the spatial distribution, sizes or clearing timescales of star-forming regions (i.e., molecular clouds), or contrasting recent star-formation histories.
\end{abstract}



\section{Introduction}
\label{sec:introduction}

Spiral arms are common features in low-redshift galaxies, with as many as two-thirds of galaxies in the low-redshift Universe exhibiting spiral structure \citep{Nair_10,Lintott_11,Willett_13}. Spiral arms are sites of enhanced gas (e.g., \citealt{Grabelsky_87,Engargiola_03}), star formation (e.g., \citealt{Calzetti_05,Grosbol_12}) and dust density \citep{Holwerda_05} compared to the interarm regions of galaxy discs. The term spiral, however, encompasses a range of galaxies with varying physical characteristics. To this end, spiral galaxies are commonly described as either grand design, many-armed or flocculent \citep{EE_82,EE_87}. Grand design spiral galaxies have two strong spiral arms propagating through the entire disc, whereas many-armed or flocculent galaxies associated with more fragmented spiral structure. In order to gain a complete understanding of the processes that link spiral arms with star formation, star formation in all types of spiral galaxy must be considered.

In the low-redshift Universe, overall star formation rates (SFRs) follow scaling relations with respect to galaxy stellar mass \citep{Brinchmann_04,Salim_07} and gas density \citep{Kennicutt_98}. The tightness of the relationship between total SFR and stellar mass indicate that the processes responsible for star formation are regulated  \citep{Bouche_10,Lilly_13,Hopkins_14}, and apply to all galaxies, irrespective of morphology. Further scaling relations between SFR density and gas density within individual galaxies \citep{Kennicutt_98,Leroy_08,Bigiel_08} and of SFR with total gas mass \citep{Saintonge_16} indicate that the current SFR of low-redshift galaxies is tied to the availability of gas to form new stars \citep{Saintonge_13,Genzel_15}, and that star formation efficiency varies little within or between galaxies \citep{Kennicutt_98,Saintonge_11}. 

Spiral arms have been linked to enhanced star formation as they are sites of increased density of young stars and gas in the Milky Way \citep{Morgan_53,McGee_64,Grabelsky_87,Elmegreen_87} and other local spiral galaxies \citep{Lada_88,Boulanger_92,Engargiola_03,Calzetti_05}.
These arms can theoretically arise in many different ways, which affect the star formation properties of galaxies. Spiral density waves are a candidate mechanism for the formation of two-armed spiral structure, and were suggested to trigger star formation where in the neighbourhoods of individual arms \citep{Lindblad_63,Lin_64,Roberts_69}. However, there is little evidence for the triggering of star formation globally in galaxies by spiral arms \citep{Romanishin_85,EE_86,Stark_87}, or within the arms of individual local galaxies \citep{Foyle_11,Dobbs_11,Eden_12,Choi_15}.  Alternatively, grand design spiral patterns could arise when the Toomre Q value in discs reaches $\sim$1, and be subject to swing amplification \citep{Toomre_64,Toomre_81}, be remnants of recent tidal interactions \citep{Sundelius_87,Dobbs_10}, or form via bar instabilities \citep{Kormendy_79}. Many-armed or flocculent spiral patterns, however, form via different mechanisms, and are more transient, short-lived structures in gas rich discs (e.g., \citealt{Sellwood_84,Baba_13,Donghia_13}). Given the little evidence for triggering of star formation by any of these mechanisms, spiral arms appear to concentrate the star-forming material into the arm regions. Star formation reflects the distribution of gas, but the arms do not affect the overall star formation in the host galaxy \citep{Vogel_88,Elmegreen_02,Moore_12}.

In this paper, the star formation and gas properties of spiral galaxies are compared with respect to spiral arm number. We use the visual classifications from Galaxy Zoo 2 (GZ2; \citealt{Willett_13}) to define samples of spiral galaxies differentiated by arm number \citep{Hart_16}. These are compared by combining estimates of SFRs measuring unobscured ultraviolet (UV) emission and obscured mid-infrared (MIR) emission. Atomic gas fractions are also compared to investigate whether the presence of different types of spiral structure lead to deviations in star formation efficiency. 

This paper is organised as follows. In Section 2, the sample selection and galaxy data are described. In Section 3, the SFR and gas properties of galaxies with different numbers of spiral arms are compared. The results and their implications with respect to relevant theoretical and observational literature are discussed in Section 5. The results are summarised in Section 5.

This paper assumes a flat cosmology with $\Omega_\mathrm{m}=0.3$ and $H_0=70 \mathrm{km s^{-1} \, Mpc^{-1}}$.

\section{Data}
\label{sec:data}

\subsection{Galaxy properties}
\label{sec:galaxy_properties}

All galaxy morphological information is obtained from the public data release of GZ2\footnote{https://www.zooniverse.org/} \citep{Willett_13}. As this paper concerns the detailed visual morphologies of spiral galaxies, we make use of the updated visual classifications given in \citet{Hart_16}, which are designed to give more consistent classifications for the multiple-answer questions in GZ2, such as spiral arm number.\footnote{The GZ2 classifications are available from http://data.galaxyzoo.org/.} The galaxies classified in GZ2 were taken from the SDSS main galaxy sample, which is an $r$-band selected sample of galaxies in the legacy imaging area targeted for spectroscopic follow-up \citep{Strauss_02}. The \citet{Hart_16} sample contains all well-resolved galaxies in SDSS DR7 \citep{Abazijian_09} to a limiting magnitude of $m_r \leq 17.0$. In this paper, we consider galaxies classified in the normal-depth (single-epoch) DR7 imaging with spectroscopic redshifts. Spectroscopic redshifts are required for galaxies to have accurate morphological data corrected for redshift-dependent classification bias (see \citealt{Bamford_09} and \citealt{Hart_16}), to allow for accurate $1/V_\mathrm{max}$ corrections to the data (see Sec.~\ref{sec:sample_selection}), and for accurate measurements of rest frame photometry.

Rest-frame absolute $ugriz$ magnitudes for are obtained from the NASA Sloan Atlas (NSA; \citealt{Blanton_11}). This restricts our sample to redshifts below 0.055, which also ensures our morphological information is robust. A low-redshift limit of $z \geq 0.02$ is also applied to remove any galaxies with large angular sizes, which may have associated morphological, spectroscopic and photometric complications. In total, there are 62,903 NSA galaxies in the redshift range $0.02 \leq z \leq 0.055$ which were visually classified in GZ2. 

In order to study the star formation properties of the galaxies in the sample, photometric data in the UV and IR are required. UV absolute magnitudes are obtained from the GALEX GR6 catalogue \citep{Martin_05}, which are also included in the NSA. Near-IR (NIR) and mid-IR (MIR) photometry are from the AllWISE catalogue of galaxies from the WISE mission \citep{Wright_10}, and obtained from the reduced catalogue of \citet{Chang_15}. We only match WISE detections to galaxies where there is only one WISE source within 6\arcsec{} of a galaxy, in line with \citet{Donoso_12}, \citet{Yan_13} and \citet{Chang_15}, and with minimum SNR>2 in at least one of the WISE bands. 45,192 (71.8 per cent) of the NSA galaxies with measured morphologies have unambiguous WISE matches with reliable photometry. We refer to this selection as the \textit{full sample}. Galaxy stellar masses are obtained from the SDSS-WISE SED fitting of \citet{Chang_15} for all galaxies in the \textit{full sample}.

To investigate the gas properties of the galaxies in the \textit{full sample}, we use measured gas masses from the $\alpha 70$ data release of the ALFALFA survey \citep{Giovanelli_05,Haynes_11}. We select reliable HI detections using objects with ALFALFA detcode=1 or 2 (described in \citealt{Haynes_11}) and a single SDSS matched optical counterpart in our  \textit{full sample} within the redshift range $0.02 \leq z \leq 0.05$. Due to the restrictions on the $\alpha 70$-SDSS footprint and the imposed limiting redshift of $z \leq 0.05$, 20,024 galaxies from the \textit{full sample} are targeted by ALFALFA and 5,570 of those galaxies have reliable H\textsc{i} fluxes.



\subsection{Sample selection}
\label{sec:sample_selection}

\begin{figure}
    \includegraphics[width=0.45\textwidth]{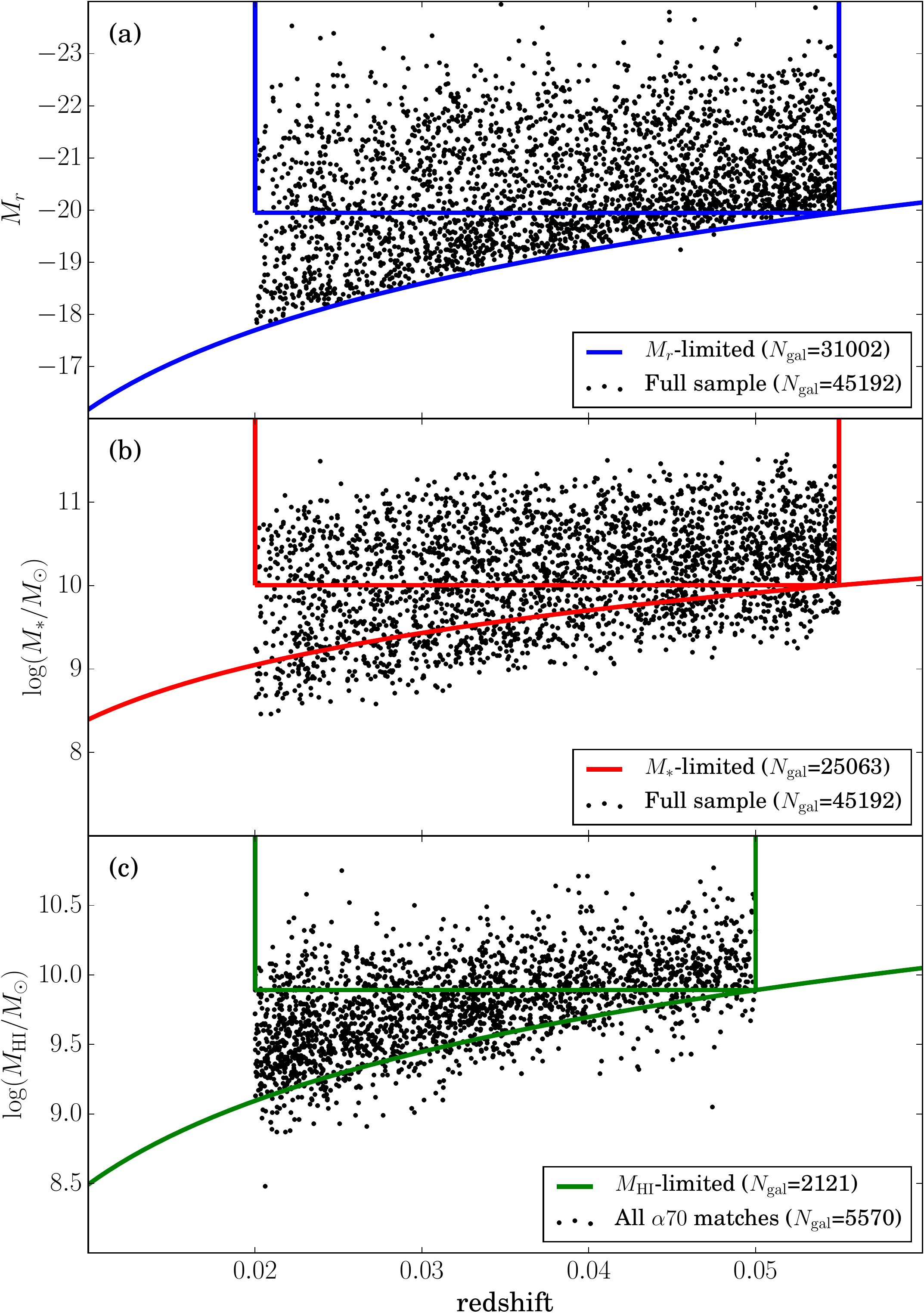}
    \caption{(a) Plot of absolute magnitude vs. redshift for the \textit{full sample} of galaxies. The curved blue line indicates the luminosity limit as a function of redshift. Galaxies enclosed within the blue box make up the \textit{luminosity-limited sample}. (b) Stellar mass distribution of the \textit{full sample} vs. redshift. The curved line shows the calculated stellar mass completeness limit and galaxies. Galaxies in the red boxed region are included in the \textit{stellar mass-limited sample}. (c) Gas mass vs. redshift for all galaxies matched in $\alpha$70 to our \textit{full sample}. The curved line shows the calculated H\textsc{i} mass completeness limit and galaxies. Galaxies in the green boxed region are included in the \textit{H\textsc{I} mass-limited sample}.}
    \label{fig:samples}
\end{figure}

The GZ2 parent sample has an apparent magnitude limit of $m_r < 17.0$. The corresponding limit in absolute magnitude at $z=0.055$ is $M_r = -19.95$. This \textit{luminosity-limited sample} is indicated by the blue box of Fig.~\ref{fig:samples}(a). The limit above which the sample is complete for a given stellar mass changes as a function of redshift according to:
\begin{equation}
	\label{eq:mass_completeness_line}
    \log(M_{*\mathrm{,lim}}) = 2.17\log(z) + 12.74 \;, 
\end{equation}
which is indicated by the curved line of Fig.~\ref{fig:samples}(b). The sample is still incomplete for the reddest galaxies at $\log(M_*/M_\odot) < 10.0$. We therefore define a \textit{stellar mass-limited sample} of galaxies, which includes all galaxies with $M_r \leq -19.95$ and $\log(M_*/M_\odot) \geq 10.0$. The limits of the sample are indicated by the red box region in Fig.~\ref{fig:samples}(b), and it includes 25,063 galaxies in total.

Similar completeness limits apply to the ALFALFA data: at a given redshift, the sample is incomplete for the least luminous H\textsc{i} sources. 
For a source of profile width 200 $\mathrm{kms^{-1}}$ ALFALFA has a 5$\sigma$ completeness limit of $S_\mathrm{lim} \geq 0.72$ \citep{Giovanelli_05}, where $S$ is the H\textsc{i} flux density. ALFALFA fluxes are converted to gas masses using the following equation \citep{Giovanelli_05}: \begin{equation}
	\label{eq:gas_mass}
	M_\mathrm{H\textsc{i}} = 2.356\times 10^5 D_\mathrm{Mpc}^2 S_\mathrm{Jy km s^{-1}} \; ,
\end{equation} and the estimated completeness limit at a given distance can therefore be described by: \begin{equation}
	\label{eq:gas_completeness}
	\log(M_\mathrm{H\textsc{i},lim}) = 0.72 \times (2.356\times 10^5 D_\mathrm{Mpc}^2) \; . 
\end{equation}

The limiting H\textsc{i} mass with redshift is shown by the curved green line of Fig.~\ref{fig:samples}(c). As many of the galaxies in $\alpha$70 are targeted, yet undetected, an H\textsc{i} upper limit can be measured for a galaxy at given distance using Eq.~\ref{eq:gas_completeness}.

Having defined the galaxy samples, a set of spiral galaxies are selected using the visual statistics of GZ2. Galaxies with $p_\mathrm{features \, or \, disc} \times p_\mathrm{not \, edge \, on} \times p_\mathrm{spiral} > 0.5$ and $N_\mathrm{spiral} \geq 5$ are selected in accordance with \citet{Hart_16}. In this paper, we wish to test how star formation properties vary with respect to the spiral structure rather than any other morphological differences. We therefore also exclude any strongly barred spiral galaxies, which have $p_\mathrm{bar} > 0.5$. \citet{Masters_10} used a similar cut to identify strongly barred galaxies in GZ2. An axial ratio cut of $(b/a)_g > 0.4$ is also imposed, where $a$ and $b$ are the SDSS DR7 $g$-band isophotal minor and major axis radii. This selection is used to ensure we only select face on galaxies, to avoid contamination of mis-classified galaxies and to limit the amount of reddening due to inclination. This was the same cut used in \citet{Masters_10} to identify reliable bar structures in discs. We have verified that our arm number vote fractions are consistent with inclination above this threshold. Each galaxy is then assigned a spiral arm number, depending on which of the responses to the arm number question in GZ2 had the highest vote fraction. The number of galaxies for each of the \textit{arm number subsamples} that are included in the \textit{full sample}, \textit{stellar mass-limited sample} and with H\textsc{i} detections are given in Table~\ref{table:sample_sizes}.

\begin{figure}
	
    \includegraphics[width=0.45\textwidth]{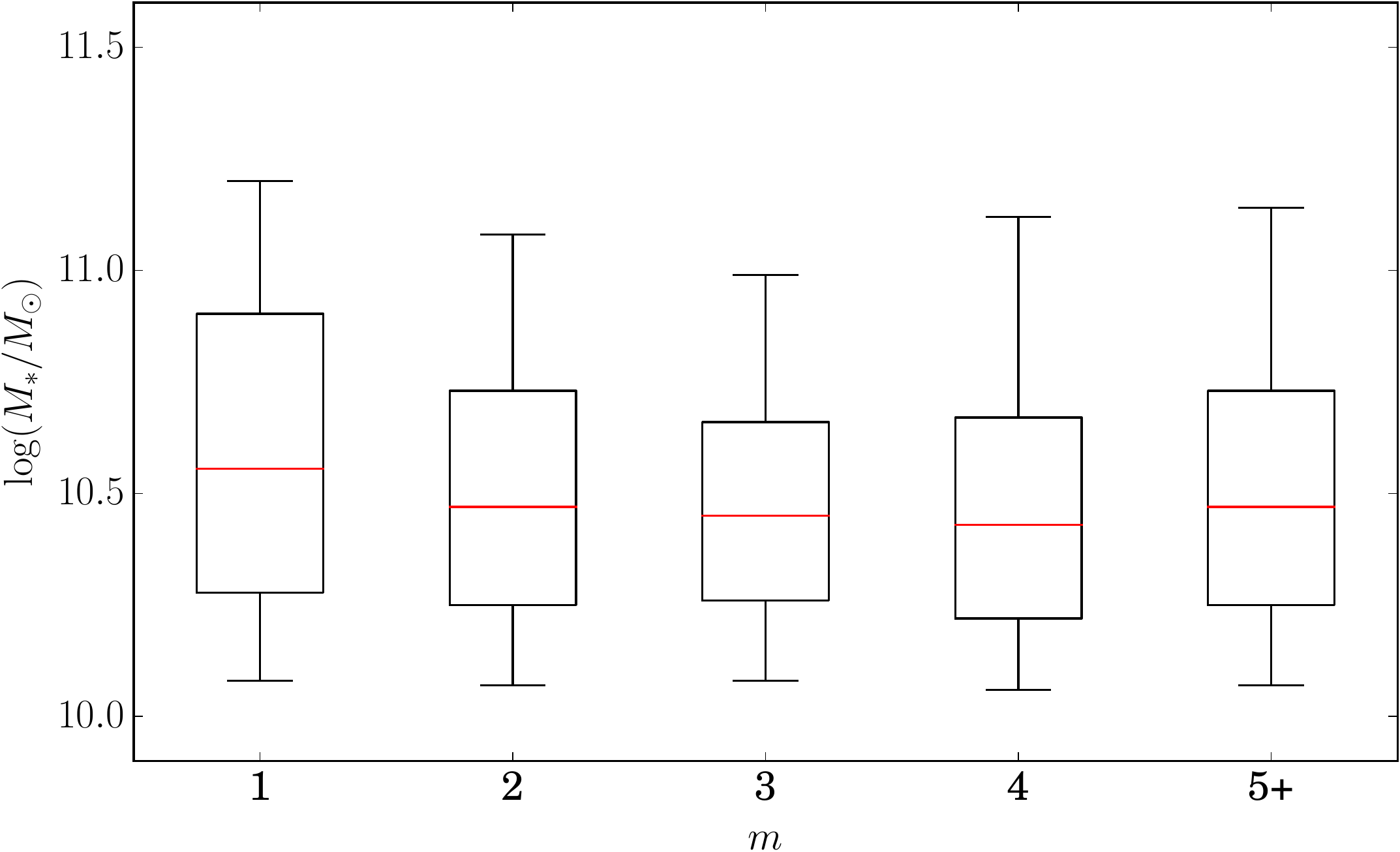}
    \caption{Stellar mass distributions (using the measurements of \citealt{Chang_15}) for each of the \textit{arm number subsamples} from the \textit{stellar mass limited sample}. The boxes show the 25th quartile, 75th quartile and the mean, and the vertical lines indicate the extent of the 5th and 95th percentiles.}
    \label{fig:mass_plot}
\end{figure}

In this paper, we make use of the \textit{stellar mass-limited sample} to compare samples of galaxies with different spiral arm numbers. As galaxy SFR is related to total stellar mass (e.g., \citealt{Brinchmann_04,Salim_07,Guo_13}), our samples must be consistent in total stellar mass to ensure that any differences in star formation properties are due to the morphological properties studied in this paper. A boxplot for each of the stellar mass distributions is shown in Figure.~\ref{fig:mass_plot}. Here we see that the stellar mass distributions of each of the samples is consistent. The $m=2$ sample has median stellar mass of $10^{10.50}M_\odot$, whereas the $m=3,4$ and $5+$ samples have medians of $10^{10.46}M_\odot$, $10^{10.44}M_\odot$ and $10^{10.49}M_\odot$. The only sample with a significantly higher median stellar mass is the $m=1$ sample, where the corresponding value is $10^{10.60}M_\odot$. However, this is the sample with the fewest galaxies (224), and the difference is still much less than the overall spread in the data (the 84th-16th percentile range for all galaxies in the \textit{stellar mass-limited sample} is $\sim 0.7$\,dex). We therefore elect to keep all galaxies in the \textit{stellar mass-limited sample} of spirals, as there is no significant stellar mass dependence on spiral arm number.
\begin{table}
\begin{tabular}{cccc}
\hline
 Morphology   &   Full sample &   $M_*$-limited &  $\alpha 70$ detected \\
\hline
 All         &         45192 &           25063 &                  5570 \\
 Spiral       &          6333 &            3889 &                  1792 \\
 $m=1$        &           482 &             224 &                   106 \\
 $m=2$        &          3298 &            1953 &                   859 \\
 $m=3$        &          1263 &             805 &                   391 \\
 $m=4$        &           534 &             357 &                   165 \\
 $m=5+$       &           756 &             550 &                   271 \\
\hline
\end{tabular}

\caption{Sample sizes for each of the samples defined in Sec.~\ref{sec:sample_selection}.}
\label{table:sample_sizes}
\end{table}

\subsection{Star formation rates}
\label{sec:sfrs}

H$\alpha$ derived SFRs are obtained from MPA-JHU measurements of SDSS spectra \citep{Brinchmann_04,Salim_07}, corrected for absorption using the Balmer decrement and for aperture effects using estimates derived from photometric galaxy colour gradients. Reliable H$\alpha$ measurements rely on spectra averaged across entire galaxies, which are not available from SDSS data alone (the SDSS fibre size is 3\arcsec{} in diameter; the median $r$-band Petrosian aperture diameter of galaxies in our \textit{stellar mass-limited sample} is 7.6\arcsec). In order to account for this, the MPA-JHU catalogue applies a correction to the fibre measured H$\alpha$ flux using photometry measured outside the fibre \citep{Salim_07}, and thus provide reliable total SFRs of star-forming galaxies \citep{Salim_16}.

Alternatively, one can measure the SFRs of galaxies using galaxy photometry rather than spectra. The UV and the MIR are usually the wavelength ranges of choice, as they are both dominated by emission from bright, young stars. The UV continuum is almost completely flat \citep{Kennicutt_98}, and arises from the direct photometric emission of the youngest stellar population. We use the conversion factor of \citet{Buat_08,Buat_11} to measure unobscured SFRs: \begin{equation}
	\label{eq:sfr_fuv}
    SFR_\mathrm{FUV} = 10^{-9.69}(L_\mathrm{FUV}/L_\odot) \; . 
\end{equation}
In order to get a reliable measure of SFR, however, the amount of UV emission that is obscured by dust must be corrected for. As dust absorbs UV photons and re-emits the energy at longer wavelengths, then one can also measure the SFR using the mid-infrared (MIR) emission. We use the following prescription from \citet{Jarrett_13}: \begin{equation}
	\label{eq:sfr_22}
    SFR_\mathrm{22} = (1-\eta) 10^{-9.125}( L_\mathrm{22}/L_\odot) \; ,
\end{equation}
where $L_\mathrm{22}$ is the luminosity measured in the WISE $22\,\mathrm{\mu m}$ band, and $\eta$ is the fraction of MIR emission that originates from the absorption of radiation from the older stellar population. Here, we use $\eta=0.17$ measured in \citet{Buat_11}. This conversion is for a \citet{Kroupa_01} IMF, which we convert to a \citet{Chabrier_03} IMF by adjusting the SFR by $-0.03$\,dex as suggested in \citet{Zahid_12} and \citet{Speagle_14}. To reliably measure the total SFR, the $22\,\mathrm{\mu m}$-derived SFR is then be added to the unobscured FUV measured SFR (e.g., \citealt{Wang_96,Heckman_98,Hao_11,Jarrett_13,Clark_15}): \begin{equation}
	\label{eq:sfr_total}
    SFR_\mathrm{total} = SFR_\mathrm{FUV} + SFR_\mathrm{22} \; . 
\end{equation} The specific star formation rate is given by $sSFR=SFR_\mathrm{total}/M_*$.

\begin{figure}
    \includegraphics[width=0.45\textwidth]{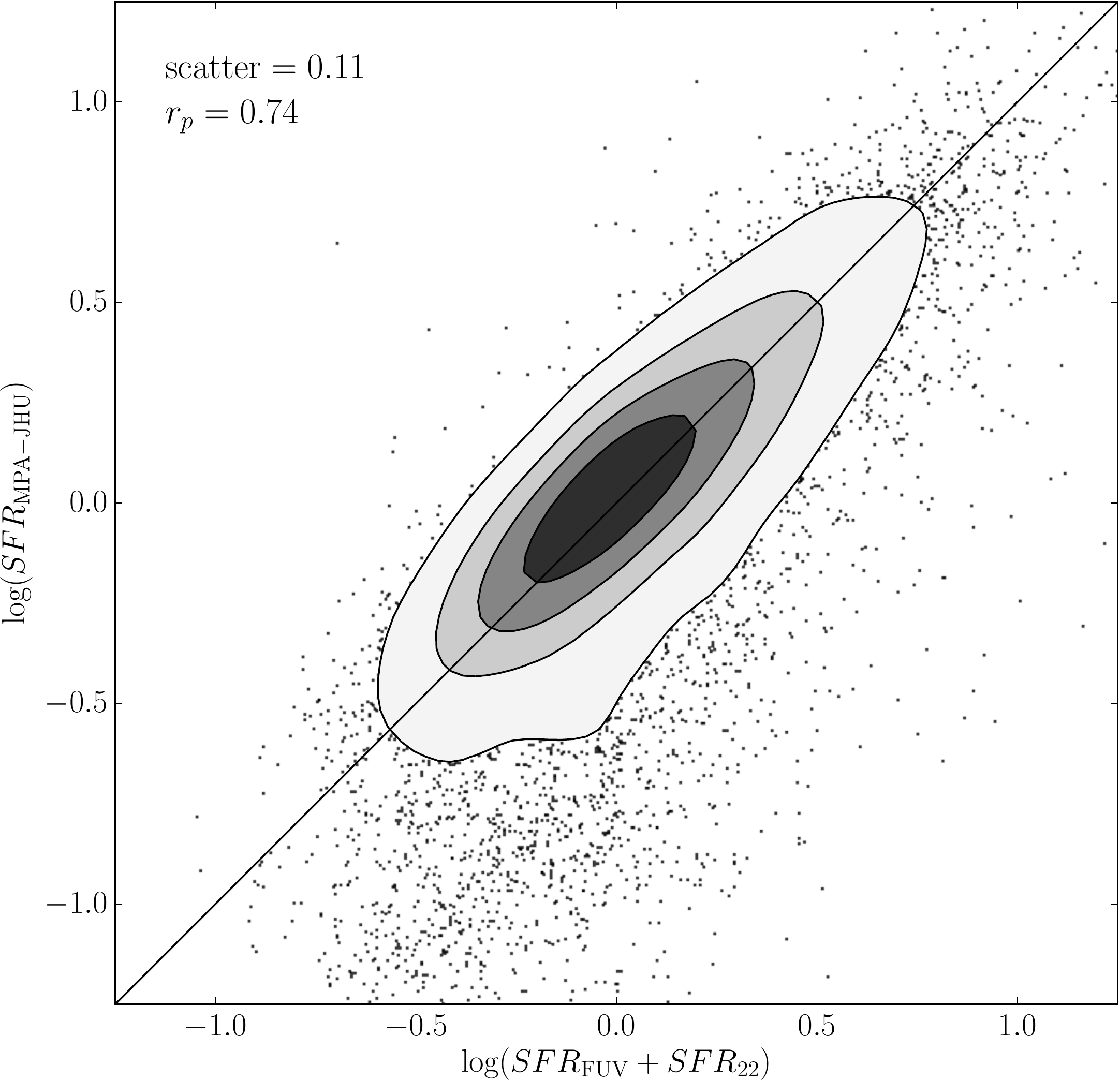}
    \caption{SFRs from Eq.~\ref{eq:sfr_total} compared to the values measured from the MPA-JHU catalogue \citep{Brinchmann_04} for galaxies in the redshift range $0.02 < z \leq 0.055$ with $\mathrm{SNR} > 2$ in the GALEX FUV and WISE $22\,\mathrm{\mu m}$ bands. The shaded grey contours show the regions enclosing 20, 40, 60 and 80\% of the data points, and the thinner black line shows the expected one-to-one relationship.}
    \label{fig:sfr_measures}
\end{figure}

To check the reliability of the SFRs obtained from this measure and to ensure our results are consistent with \citet{Willett_15}, we compare the SFRs from Eq.~\ref{eq:sfr_total} with the MPA-JHU H$\alpha$ estimates in Fig.~\ref{fig:sfr_measures}. The plot shows all galaxies in the redshift range $0.02 < z \leq 0.055$ which have $\mathrm{SNR} > 2$ in both the GALEX FUV and the WISE $22\,\mathrm{\mu m}$. The GALEX FUV band is complete for galaxies with $L_\mathrm{FUV} \geq 8.5 L_\odot$ at $z=0.02$ and $L_\mathrm{FUV} \geq 8.9 L_\odot$ at $z=0.055$. The WISE $22\,\mathrm{\mu m}$ band is complete for galaxies with $L_\mathrm{FUV} \geq 7.5 L_\odot$ at $z=0.02$ and $L_\mathrm{FUV} \geq 8.4 L_\odot$ at $z=0.055$.  There is good agreement between the FUV+MIR derived SFRs and those inferred from the Balmer lines, with Pearson's r coefficient of $0.74$, indicative of a strong linear correlation between the variables. The scatter is $0.11$\,dex. There is also no significant offset between the measurements (the median difference is $< 0.01$\,dex), suggesting that these SFR measurements are indeed comparable. We do however see galaxies at the lower end of $SFR_\mathrm{MPA-JHU}$ with higher SFRs measured from Eq.~\ref{eq:sfr_total}. \citet{Salim_16} attributes galaxies with higher SFRs measured in \citet{Chang_15} to galaxies with low measured fluxes in WISE being overestimated. However, we do not expect this effect to dominate as we find strong agreement between the SFR indicators using a single value of $\eta = 0.17$ from \citet{Buat_08}, with the only differences observed for galaxies with low SFRs ($\lesssim 0.5 \mathrm{M_\odot yr^{-1}}$). This issue affects only a small fraction of galaxies in the sample, with $< 10$ per cent of galaxies having more than $0.5$\,dex disagreement between the two SFR measures. We therefore elect to keep all galaxies with $\mathrm{SNR} > 2$ in both the GALEX FUV and the WISE $22\,\mathrm{\mu m}$ in the sample of galaxies with reliably measured SFRs.

\section{Results}
\label{sec:results}

In this section, we use the galaxy SFRs described in Sec.~\ref{sec:sfrs} to investigate how the galaxy star formation properties of our GZ2 galaxies vary with respect to spiral arm number. The FUV+MIR SFR measurements give a measure of SFR obtained from across the entire galaxy, and are thus sensitive to any SFR differences in the discs of galaxies. They also give an opportunity to assess the relative fractions of obscured and unobscured star formation directly, which will be investigated in Sec.~\ref{sec:uv_vs_mir}.

\subsection{The Star Formation Main Sequence}
\label{sec:sfms}

The star formation main sequence (SFMS) describes the SFR of the galaxy population as a function of stellar mass. In the low-redshift Universe, this correlation has been shown to be very tight for normal star-forming galaxies \citep{Brinchmann_04,Salim_07,Chang_15}. Galaxies with significantly enhanced star formation are usually associated with merging or interacting systems \citep{Sanders_96,Veilleux_02,Engel_10,Kaviraj_14,Willett_15}, with the rest of the difference across the main sequence attributable to differences in the gas content and star formation efficiency of galaxies \citep{Saintonge_11,Saintonge_16}. 

\begin{figure}
    \includegraphics[width=0.45\textwidth]{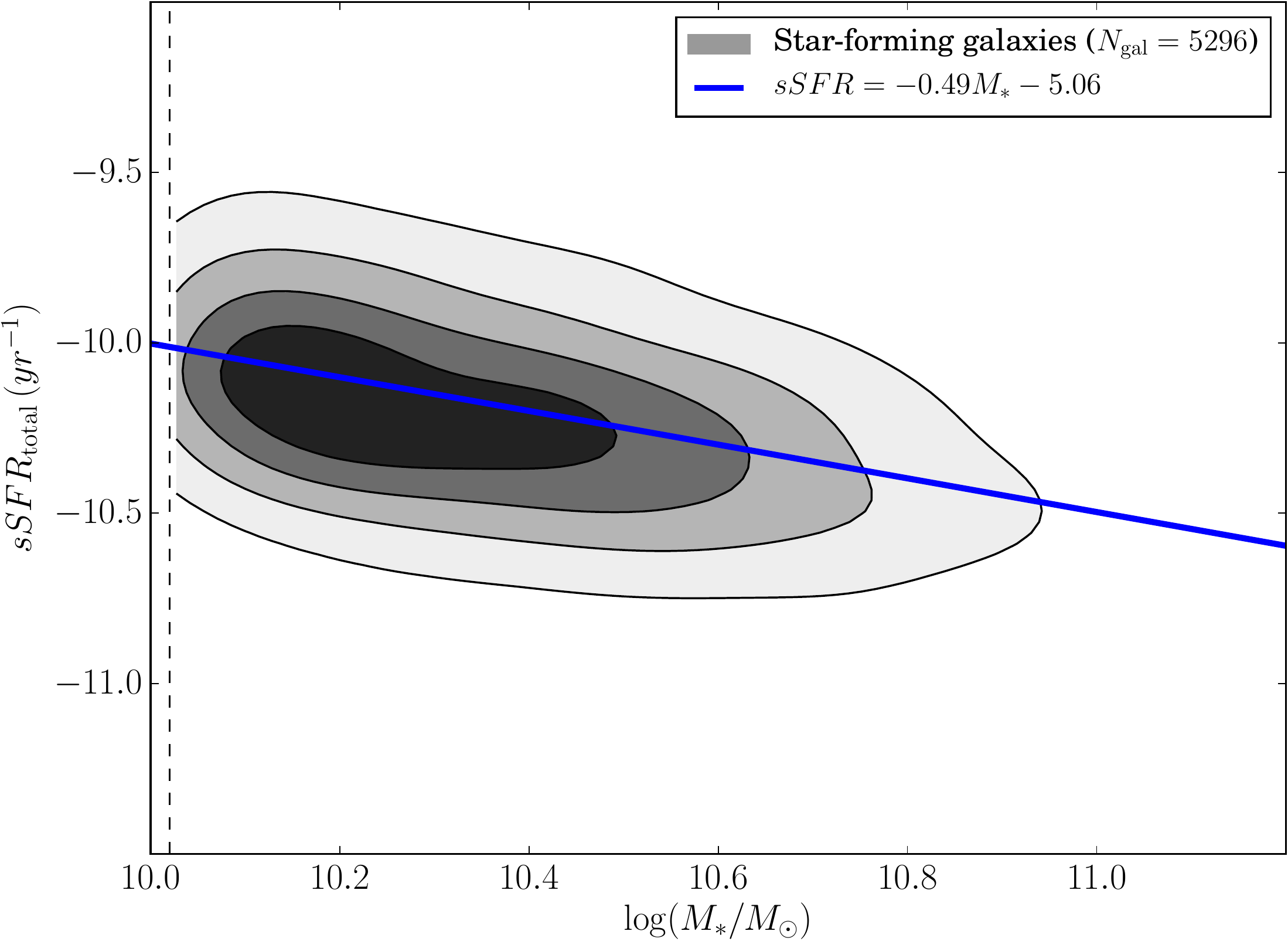}
    \caption{Stellar mass vs. sSFR for star-forming galaxies in the redshift range $0.02 < z \leq 0.05$ The grey contours show regions enclosing 20, 40, 60 and 80 per cent of the points. The blue line indicates the linear best fit line to the data. The black dashed vertical line indicates the galaxy of lowest stellar mass in this sample.}
    \label{fig:sfms_all}
\end{figure}

In order to test whether the morphology of any of our galaxies affect the total star formation rate, the SFMS is plotted using the definition of sSFR defined in Sec.~\ref{sec:sfrs}. We elect to plot sSFR rather than SFR, as this more clearly demonstrates how efficiently gas is converted to stars in star-forming galaxies with respect to stellar mass, $M_*$. Galaxies at low-redshift can be considered bimodal in terms of their colour and SFR properties \citep{Baldry_06,Schawinski_14,Morselli_16}. To plot the SFMS, we must therefore first select a set of galaxies that are considered star-forming. We choose to use the definition of \citet{Chang_15}, which defines star-forming galaxies using SDSS \textit{ugriz} photometry. Using this definition, we select galaxies with $(u-r)_\mathrm{rest} < 2.1$ or $(u-r)_\mathrm{rest} < 1.6(r-z)_\mathrm{rest} + 1.1$ as star-forming. The majority of spiral galaxies (78.3 per cent) are found to be star-forming using this method. The resulting plot of $\log(M_*)$ vs. $\log(sSFR)$ is shown in Fig.~\ref{fig:sfms_all}. As expected, we see a tight relationship, as galaxies with greater stellar masses have lower sSFRs. The best-fit linear model to the data is given by:\begin{equation}
	\label{eq:SFR_expected}
    \log(sSFR_\mathrm{expected}) = -0.49\log(M_*/M_\mathrm{\odot}) - 5.06 \; ,
\end{equation}
and the scatter is $0.19$\,dex. This relationship can now be used to assess whether galaxies have systematically high or low sSFRs for their given $M_*$. We do this by defining the best-fit line as the \textit{expected} sSFR for a galaxy of a given stellar mass. Given this information, we define the $sSFR_\mathrm{residual}$ in Eq.~\ref{eq:SFR_residual}:
\begin{equation}
	\label{eq:SFR_residual}
    \log(sSFR_\mathrm{residual}) = \log(sSFR_\mathrm{total}) - \log(sSFR_\mathrm{expected}) \; ,
\end{equation}
where $\log(sSFR_\mathrm{expected})$ is given in Eq.~\ref{eq:SFR_expected}. If a given galaxy lies above the $sSFR_\mathrm{expected}$ line, it has a higher sSFR for its stellar mass compared to the total star-forming galaxy population, and a positive value for $sSFR_\mathrm{residual}$. Conversely, galaxies below the line can be considered as being deficient in sSFR, and have a negative value for $sSFR_\mathrm{residual}$. 

Using equations \ref{eq:SFR_expected} and \ref{eq:SFR_residual}, the effect that spiral galaxy morphology has on the total sSFRs of galaxies is now considered. For this analysis, we use a subsample of spiral galaxies, which we split into \textit{arm number subsamples}, using the morphology criteria described in Sec.~\ref{sec:sample_selection}. We consider only galaxies taken from the \textit{stellar mass-limited sample}, as these galaxies are well matched in stellar mass (see Sec.~\ref{sec:sample_selection}).

\begin{figure}
	
    \includegraphics[width=0.45\textwidth]{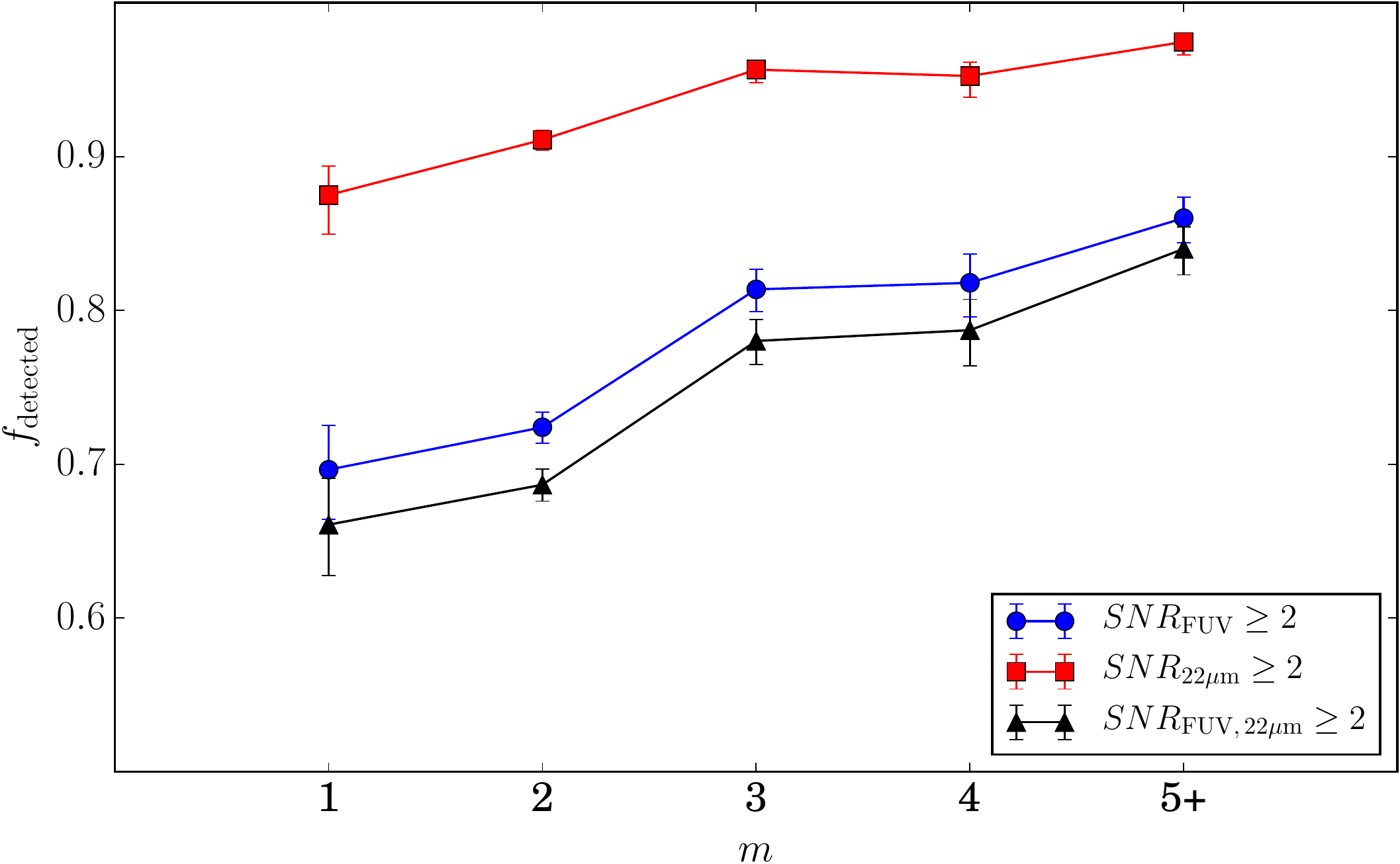}
    \caption{Fraction of galaxies with $\mathrm{SNR} > 2$ detection in the GALEX FUV (blue circles), WISE $22\,\mathrm{\mu m}$ (red squares), and both (black triangles), for each of the \textit{arm number subsamples} taken from the \textit{stellar mass-limited} sample of spiral galaxies. The errorbars show the $1\sigma$ errors, calculated using the method of \citet{Cameron_11}.}
    \label{fig:sfr_completeness}
\end{figure}

As discussed in Sec.~\ref{sec:sfrs}, we impose cuts in SNR to ensure that we have flux measurements that are not dominated by noise to get a reliable estimate of SFR in both the FUV and the MIR. It is therefore important to first check the completeness of each of the samples that we compare. The fraction of galaxies that meet the minimum $\mathrm{SNR} > 2$ threshold in the GALEX FUV filter, the WISE $22\,\mathrm{\mu m}$ filter, and both filters,  are shown in Fig.~\ref{fig:sfr_completeness}. The overall completeness of each of the samples is similar, with $\sim 70$--$80$ per cent of galaxies having a detection in both filters.  We do see that the many-armed samples ($m=3$, $4$ or $5+$) have a greater fraction of galaxies with reliable fluxes in both the $22\,\mathrm{\mu m}$ and the GALEX FUV than the $m=1$ and $m=2$ sample, however. Thus, galaxies with one or two spiral arms are more likely to have undetectable MIR or FUV emission and thus low SFRs. 

\begin{figure}
	
    \includegraphics[width=0.45\textwidth]{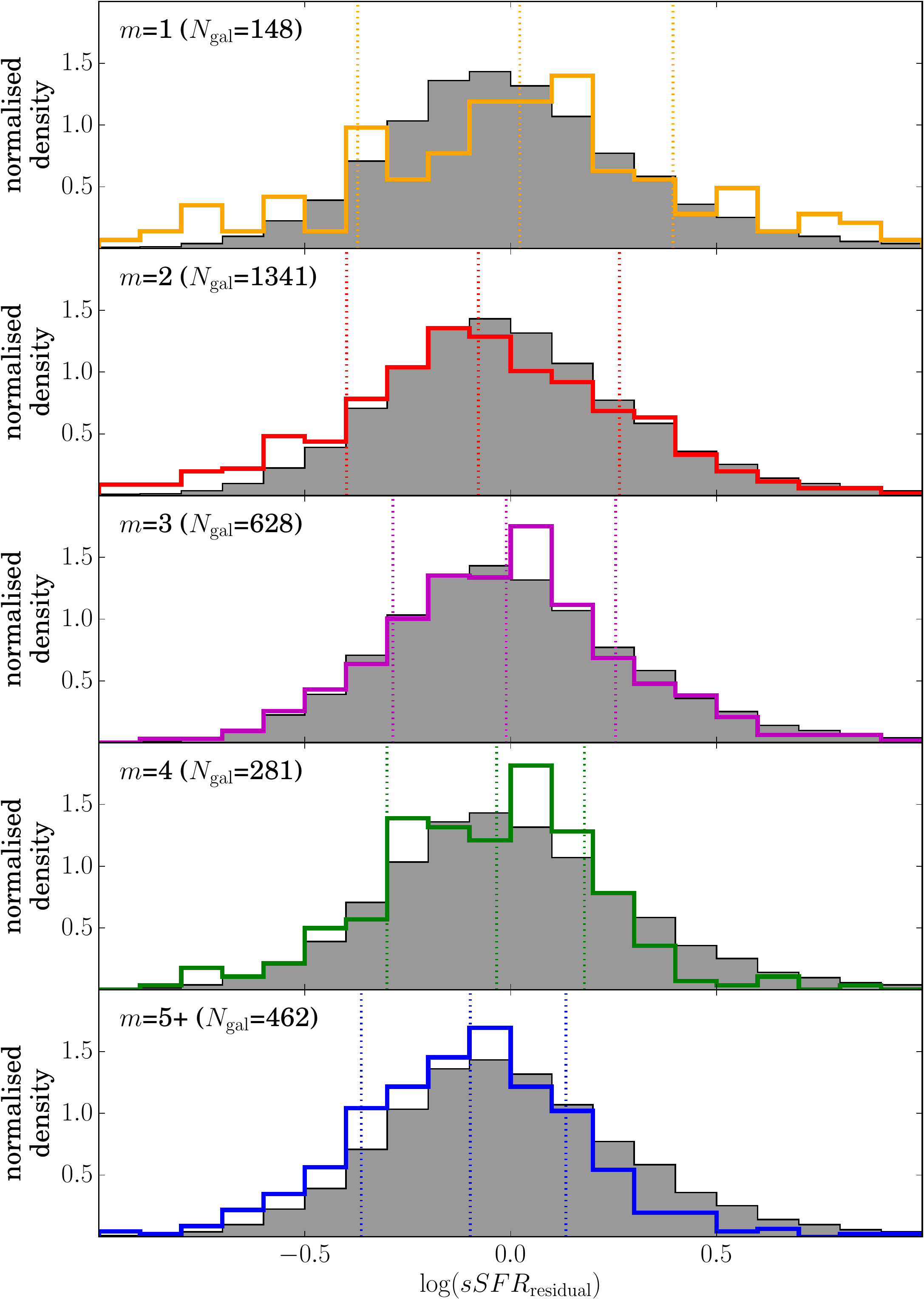}
    \caption{Residual sSFRs for each of the \textit{arm number subsamples} taken from the \textit{stellar mass-limited sample}, calculated using equations ~\ref{eq:SFR_expected} and ~\ref{eq:SFR_residual}. The solid histograms show the distributions for each subsample, and the filled grey histograms indicate the same distributions for the entire sample of \textit{star forming} galaxies for reference. The vertical dotted lines indicate the 16th, 50th and 84th percentiles.}
    \label{fig:ssfr_residual}
\end{figure}

The resulting distributions of $sSFR_\mathrm{residual}$ for each of the \textit{arm number subsamples} are shown in Fig.~\ref{fig:ssfr_residual}. Only galaxies with reliable SNR>2 measurements in both the GALEX FUV and the WISE 22$\mu \mathrm{m}$ are included in these distributions (the fractions that meet these requirements for each \textit{arm number subsample} are shown in Fig.~\ref{fig:sfr_completeness}). It is immediately apparent that there is no strong dependence of $sSFR_\mathrm{residual}$ on spiral arm number -- the median of the $m=2$ distribution compared with the $m=3$, $4$, $5+$ distributions shift by $\lesssim 0.05$\,dex, which is much smaller than the scatter in the SFMS of $0.19$\,dex. This result is perhaps surprising, given that in \citet{Hart_16} it was shown that the many-armed samples are much bluer in colour compared to their two-armed counterparts, an effect which was suggested to be related to the star formation properties of the galaxies. The $m=1$ sample of spiral galaxies has the highest median value of $sSFR_\mathrm{residual}$ of $0.02 \pm 0.05$, and is the only sample which lies above the defined SFMS, although this is \textit{arm number subsample} with the lowest number of galaxies with reliable FUV and 22$\mu \mathrm{m}$ measurements (148 galaxies). These high sSFRs are likely because GZ2 classified $m=1$ spiral galaxies are associated with tidally induced features \citep{Casteels_13}, which are in turn associated with enhanced star formation \citep{Sanders_96,Veilleux_02,Engel_10}. This result is also expected, as \citet{Willett_15} showed that spiral arm number does not affect the position of the SFMS, albeit using the spectroscopic SFRs of \citet{Brinchmann_04}. Merger systems on the other hand did show SFRs above the SFMS.  It should also be noted that each of the galaxy distributions lie very close to the SFMS: we make no cut in only selecting star-forming galaxies for this analysis, yet the medians of each of the spiral galaxy subsamples are within $\lesssim 0.1$\,dex of the SFMS. Although a significant number of galaxies are observed to be passively star-forming, \citep{Masters_10,Fraser-McKelvie_16}, this population cannot be attributed to galaxies of a specific spiral arm number -- the majority of galaxies with \textit{any} spiral arm number are actively star-forming.

\subsection{Comparing obscured and unobscured star formation}
\label{sec:uv_vs_mir}

As discussed in Sec.~\ref{sec:sfrs}, the different SFR indicators that we use to define our total SFRs in Eq.~\ref{eq:sfr_total} correspond to the combination of emission from the youngest, hottest stars (measured in the UV), and emission originating from the radiation absorbed by interstellar dust and re-emitted at longer wavelengths (the MIR). Although both sources of emission arise from young stars of order $\sim$10 Myr in age \citep{Hao_11,Rieke_09}, their relative contributions actually trace different phases in the molecular gas clouds from which they form. \citet{Calzetti_05} noted that the MIR emission from galaxies traces the H$\alpha$ emission, which itself originates from absorption of photons of the youngest stars (<$10$ Myr in age), suggesting that warm dust emission is attributable to the star-forming regions of galaxies. UV emitting populations are instead visibly offset from the most active star-forming regions \citep{Calzetti_05}. In the absence of highly star-forming starburst galaxies, the processes via which the UV population become exposed take some time, and are highly dependent on the gas and star formation conditions \citep{Parravano_03} of the molecular clouds from which stars form.

\begin{figure}
    \includegraphics[width=0.45\textwidth]{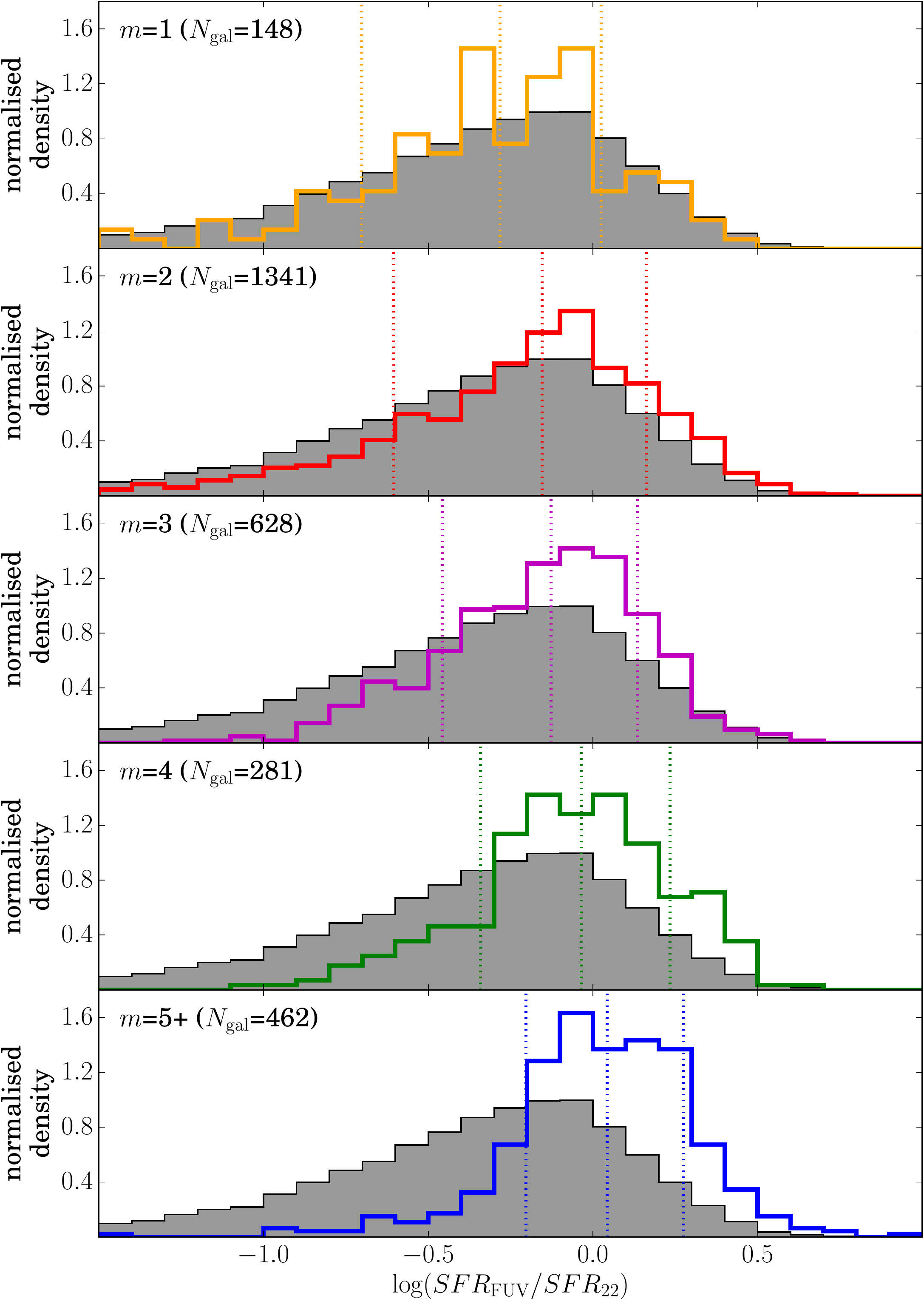}
    \caption{sSFRs measured in the FUV and MIR for the \textit{stellar mass-limited samples}. The grey filled histogram shows the same distribution for all galaxies in the \textit{stellar mass-limited sample}, irrespective of morphology. The vertical lines show the median, 16th and 84th percentiles for each of the \textit{arm number subsamples}.}
    \label{fig:fuv_vs_mir}
\end{figure}

To give an insight into the relative fractions of obscured and unobscured star formation in our galaxy samples, we compare the FUV and MIR components of $sSFR_\mathrm{total}$ in Fig.~\ref{fig:fuv_vs_mir}. We use the same \textit{stellar mass-limited sample} of spiral galaxies as used in Sec.~\ref{sec:sfms}. Each of the \textit{arm number subsample} populations lie close to the SFMS, so we expect them to be dominated by normal star-forming galaxies with little contribution from starburst populations. The amount of unobscured FUV star formation relative to the amount of obscured MIR measured star formation is shown in Fig.~\ref{fig:fuv_vs_mir} for each of our \textit{arm number subsamples}. Here, a clear trend is observed -- many-armed spiral galaxies have less obscured star formation than the $m=2$ sample of spiral galaxies. The $m=1$ sample has the highest median $\log(SFR_\mathrm{FUV}/SFR_{22})$ value of $-0.28 \pm 0.05$, corresponding to a mean of $34 \pm 1$ per cent of the $SFR_\mathrm{total}$ being measured by the young stars unobscured by dust in the FUV. The $m=2$, $3$, $4$ and $5+$ values for $\log(SFR_\mathrm{FUV}/SFR_{22})$ are $-0.15 \pm 0.01$, $-0.13 \pm 0.01$, $-0.03 \pm 0.02$ and $0.04 \pm 0.02$, corresponding to $40 \pm 1$, $42 \pm 1$, $47 \pm 1$ and $51 \pm 1$ per cent of the total star formation being measured in the FUV. All of the many-armed spiral galaxy subsamples have significantly higher fractions of their total SFR measured in the FUV than in the MIR -- the KS $p$-values are $\sim 10^{-3}$, $\sim 10^{-7}$ and $\sim 10^{-26}$ between the $m=2$ sample and the $m=3$, $4$ and $5+$ samples, respectively. The many-armed spiral samples have less obscured star formation than the two-armed sample, with a greater fraction of the SFR measured from young stars unattenuated by dust in the FUV.

\subsubsection{The IRX-$\beta$ relation}
\label{sec:irx_beta}

A common paramaterisation of the amount of dust obscuration in star-forming galaxies is through the IRX-$\beta$ relation \citep{Calzetti_94,Meurer_99,Calzetti_00}. The quantity IRX refers to the infrared excess, and corresponds to the relative fraction of MIR emission originating from warm dust, to the UV emission, from exposed young stars. The quantity IRX is defined by \citet{Boquien_12} as:
\begin{equation} \label{eq:IRX}
\mathrm{IRX} = \log(L_\mathrm{dust}/L_\mathrm{FUV}) \; .
\end{equation}
The quantity $\beta$ measures the slope in the UV continuum of galaxies, which depends on both the intrinsic UV slope, $\beta_0$, and the UV slope induced by dust reddening, and is defined in \citet{Boquien_12}:
\begin{equation}
	\label{eq:beta}
    \beta = \frac{M_\mathrm{FUV} - M_{\mathrm{NUV}}}{2.5\log(\lambda_\mathrm{FUV}/\lambda_\mathrm{NUV})} - 2 \; .
\end{equation}

For starbursting galaxies, the relationship between IRX and $\beta$ has been shown to be very tight, with galaxies with greater IRX having a greater UV slope \citep{Meurer_99,Kong_04,Overzier_11}. Quiescently star-forming galaxies, however, lie below the IRX-$\beta$ law for starbursting galaxies, and show significantly more scatter. Contributions to both the MIR and the UV from aging stellar populations, variations in the dust extinction properties or variations in star formation histories (SFHs) of star-forming regions have all been hypothesised as reasons why star-forming galaxies show this scatter \citep{Bell_02,Kong_04,Boquien_09,Boquien_12}.

\begin{figure*}
	\includegraphics[width=0.975\textwidth]{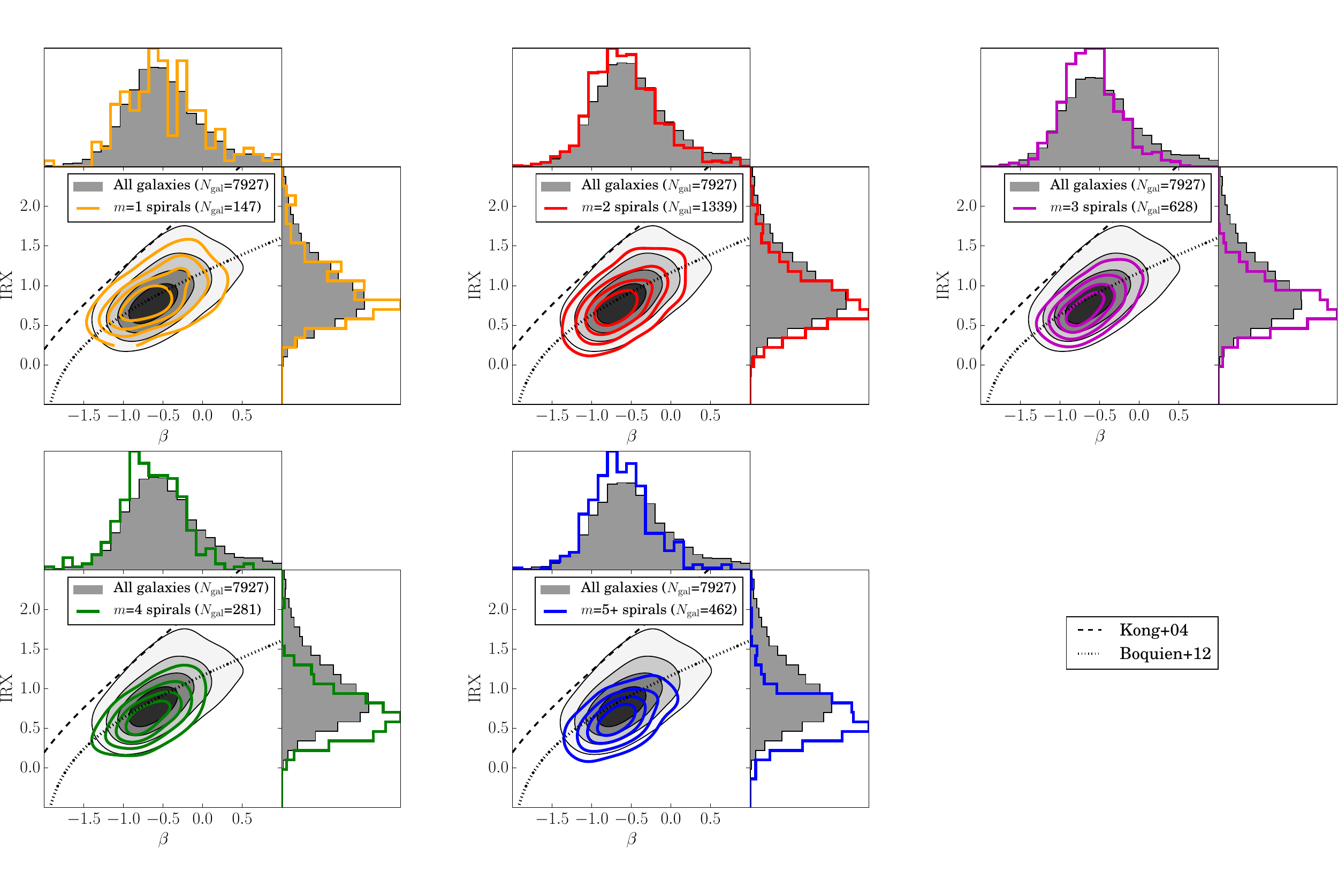}

    \caption{IRX (from Eq.~\ref{eq:IRX}) vs. $\beta$ (from Eq.~\ref{eq:beta}), for each spiral arm number. The underlying grey contours show the same distribution for all galaxies in the \textit{stellar mass-limited sample} with detections in the GALEX FUV and the WISE $22\,\mathrm{\mu m}$, regardless of morphology. The solid lines show the same values for our \textit{stellar mass-limited spiral sample}, split by spiral arm number. The black dashed line shows the IRX-$\beta$ relation measured for starburst galaxies \citep{Kong_04} and the black dotted line shows the relationship for low-redshift star-forming galaxies \citep{Boquien_12}.} 
    \label{fig:irx_beta}
\end{figure*}

We select galaxies from the \textit{stellar mass-limited sample} with $\mathrm{SNR} > 2$ detections in the GALEX FUV, GALEX NUV, and WISE $22\,\mathrm{\mu m}$, giving 7927 galaxies in total. The subset of galaxies classified as spirals using the criteria described in Sec.~\ref{sec:sample_selection} consists of 2857 galaxies. The total dust emission, $L_\mathrm{dust}$, is taken from the catalogue of \citet{Chang_15}, which fit stellar and dust emission curves to each of the galaxies. The resulting IRX-$\beta$ relation for our \textit{arm number subsamples} are shown in Fig.~\ref{fig:irx_beta}. 

All of our spiral galaxy populations lie below the IRX-$\beta$ relation from \citet{Kong_04}, with no significantly enhanced starburst-like formation. In order to measure how closely each of our samples lie to the expected IRX-$\beta$ relation of \citet{Boquien_12} (the dotted lines in Fig.~\ref{fig:irx_beta}), the median offset from the relation for a given $\beta$ is calculated; if a galaxy population lies below the relation, it has a negative median offset, and if it lies above the line, the median offset is positive. For reference, the full sample of galaxies irrespective of morphology (shown by the filled grey contours in Fig~\ref{fig:irx_beta}) has a median offset of $-0.01 \pm 0.01$, indicating that this population is representative of a normal star-forming galaxy population that follows the IRX-$\beta$ relation of \citet{Boquien_12}. The corresponding offsets for each of the \textit{arm number subsamples} are $-0.01 \pm 0.04$, $-0.05 \pm 0.05$, $-0.09 \pm 0.01$, $-0.14 \pm 0.02$ and $-0.20 \pm 0.01$. Each of the spiral galaxy populations actually lie below the IRX-$\beta$ relation, indicating that they are less luminous in the MIR than expected for their $\beta$. We also see a clear trend with spiral arm number -- the $m=1$ and $m=2$ populations lie much closer to the IRX-$\beta$ realtion for normal star-forming galaxies, whereas galaxies with more spiral arms lie further below the relation, indicating that they have more UV emission relative to MIR emission than expected for their measured $\beta$. We will discuss the implications of this further in Sec.~\ref{sec:discussion_sfr}.

\subsection{Gas properties of spiral galaxies}
\label{sec:gas}

\begin{figure}
	
    \includegraphics[width=0.45\textwidth]{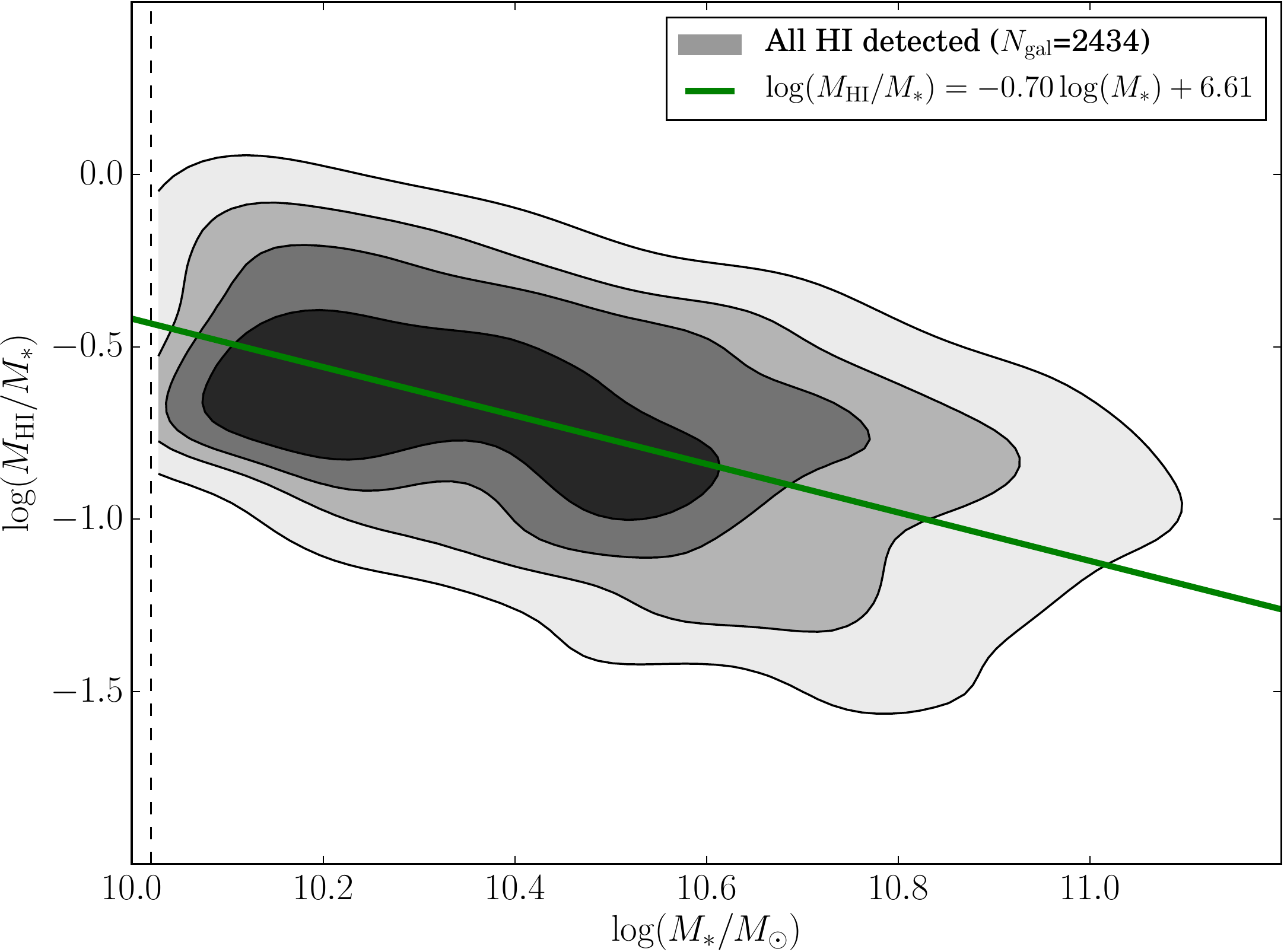}
    \caption{Gas fraction as a function of stellar mass for all galaxies in the \textit{stellar mass-limited sample} with an H\textsc{i} detection. The filled grey contours show where 20, 40, 60 and 80\% of the galaxies lie, weighted by each galaxy's $1/V_\mathrm{max}$-value. The green line shows the best fit line to the data, with each point again weighted by $1/V_\mathrm{max}$. The dashed black line indicates the lower stellar mass limit of the dataset.}
    \label{fig:gas_sequence}
\end{figure}

The amount of gas that galaxies contain is usually related to both the current star formation activity \citep{Huang_12,Saintonge_11,Saintonge_13,Saintonge_16} and galaxy morphology \citep{Helmboldt_04,Helmboldt_05,Saintonge_12,Masters_12}. However, the amount of gas in galaxy discs has little dependence on the presence of spiral structure or its type, with spiral structure instead believed to rearrange the star-forming material in galaxies \citep{Vogel_88,Elmegreen_02,Dobbs_11,Moore_12}. By using the atomic gas mass measurements of ALFALFA \citep{Giovanelli_05,Haynes_11}, we consider whether spiral structure has any link to an excess or deficiency of gas in our spiral samples. Although stars form out of molecular hydrogen, molecular clouds form out of the diffuse medium of atomic hydrogen \citep{Haynes_11}. This means that although H\textsc{i} does not directly probe the amount of star-forming material available for current star formation, it does measure the amount of material for potential star formation. The H\textsc{i} fraction, $f_\mathrm{H\textsc{i}}=M_\mathrm{H\textsc{i}}/M_{\odot}$ exhibits a strong dependence with stellar mass, with more massive galaxies having lower gas fractions \citep{Haynes_84,Cortese_11,Saintonge_16}. It is therefore useful to define the \textit{expected} gas fraction as a function of stellar mass, in order to define whether a galaxy is deficient in H\textsc{i} for its stellar mass. The value for $\log(M_\mathrm{H\textsc{i}}/M_\odot)_\mathrm{expected}$ can be calculated in a similar way to $sSFR_\mathrm{expected}$ in Sec.~\ref{sec:sfms}, by fitting a line to the plot of $\log(M_*) $ vs. $\log(M_\mathrm{H\textsc{i}}/M_*)$. The parent sample for this comparison comprises of galaxies in the \textit{stellar mass-limited sample}, including all galaxies with an H\textsc{i} detection, regardless of morphology. In order to probe the entire range of gas fractions, we use all galaxies from the \textit{stellar mass-limited sample} with a reliable H\textsc{i} detection, giving 2,434 galaxies in total. Galaxies that fall below the H\textsc{i} completeness limit with $M_\mathrm{H\textsc{i}} < 9.89$ are weighted by $1/V_\mathrm{max}$, described in Sec.~\ref{sec:sample_selection}. The plot of gas fraction vs. stellar mass for this sample is shown in Fig.~\ref{fig:gas_sequence}. The best fit line to the data, with each point weighted by $1/V_\mathrm{max}$, yields the following relationship:\begin{equation}
	\label{eq:gas_expected}
    \log(M_\mathrm{H\textsc{i}}/M_*)_\mathrm{expected} = -0.70 \log(M_*/M_\odot) + 6.61 \; . \end{equation}
The scatter in this relationship is $0.26$\,dex. One can now measure the H\textsc{i} deficiency using \citep{Masters_12}:\begin{equation}
	\label{eq:gas_deficiency}
    \log(M_\mathrm{H\textsc{i}}/M_*)_\mathrm{deficiency} = \log(M_\mathrm{HH\textsc{i}}/M_*)_\mathrm{expected} - \log(M_\mathrm{H\textsc{i}}/M_*) \; , 
\end{equation}
where $\log(M_\mathrm{H\textsc{i}}/M_*)_\mathrm{expected}$ is given in Eq.~\ref{eq:gas_expected}. Galaxies with higher gas fractions than expected for their stellar mass have negative H\textsc{i} deficiency, and galaxies with low H\textsc{i} fractions have positive H\textsc{i} deficiency. 

\begin{figure}
	
    \includegraphics[width=0.45\textwidth]{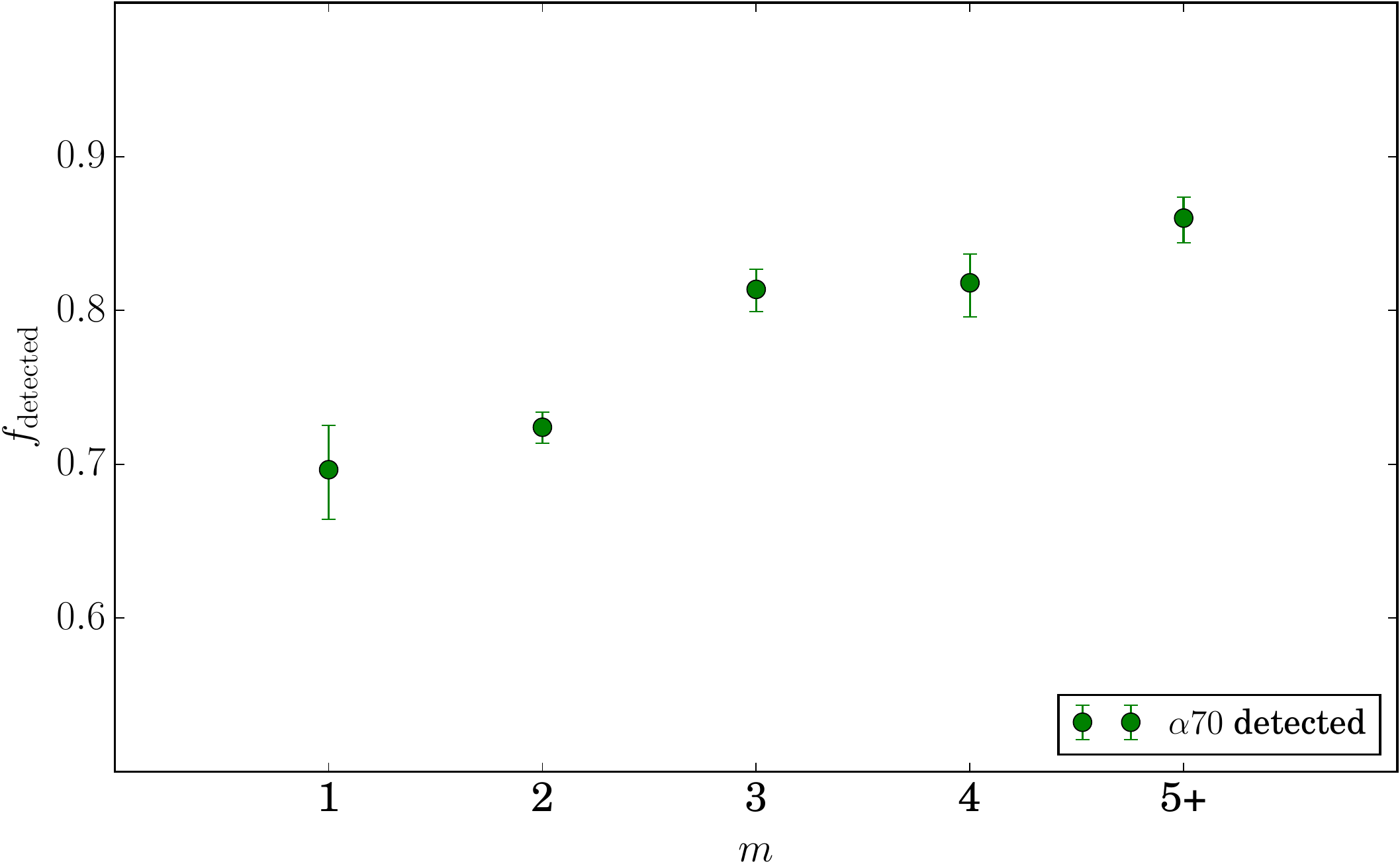}
    \caption{Fraction of galaxies in the $\alpha$70-SDSS footprint with a reliably detected H\textsc{i} flux, in accordance with \citet{Haynes_11}. The errorbars show the 1$\sigma$ error calculated in accordance with \citet{Cameron_11}.}
    \label{fig:gas_completeness}
\end{figure}

As in Sec.~\ref{sec:sfms}, we begin by comparing the completeness of our \textit{arm number subsamples}. The fraction of the \textit{stellar mass-limited sample} of ALFALFA targeted spiral galaxies with a single detection in the $\alpha$70 catalogue as a function of arm number is shown in Fig.~\ref{fig:gas_completeness}. As was the case in for the FUV and MIR fluxes, we do see a preference for more of the many-armed galaxies to have measured H\textsc{i} fluxes. However, we note that as in Sec.~\ref{sec:sfms}, the overall completeness is similar, with each of the samples having detection fractions of $\sim$70-80\%.

\begin{figure}
	
    \includegraphics[width=0.45\textwidth]{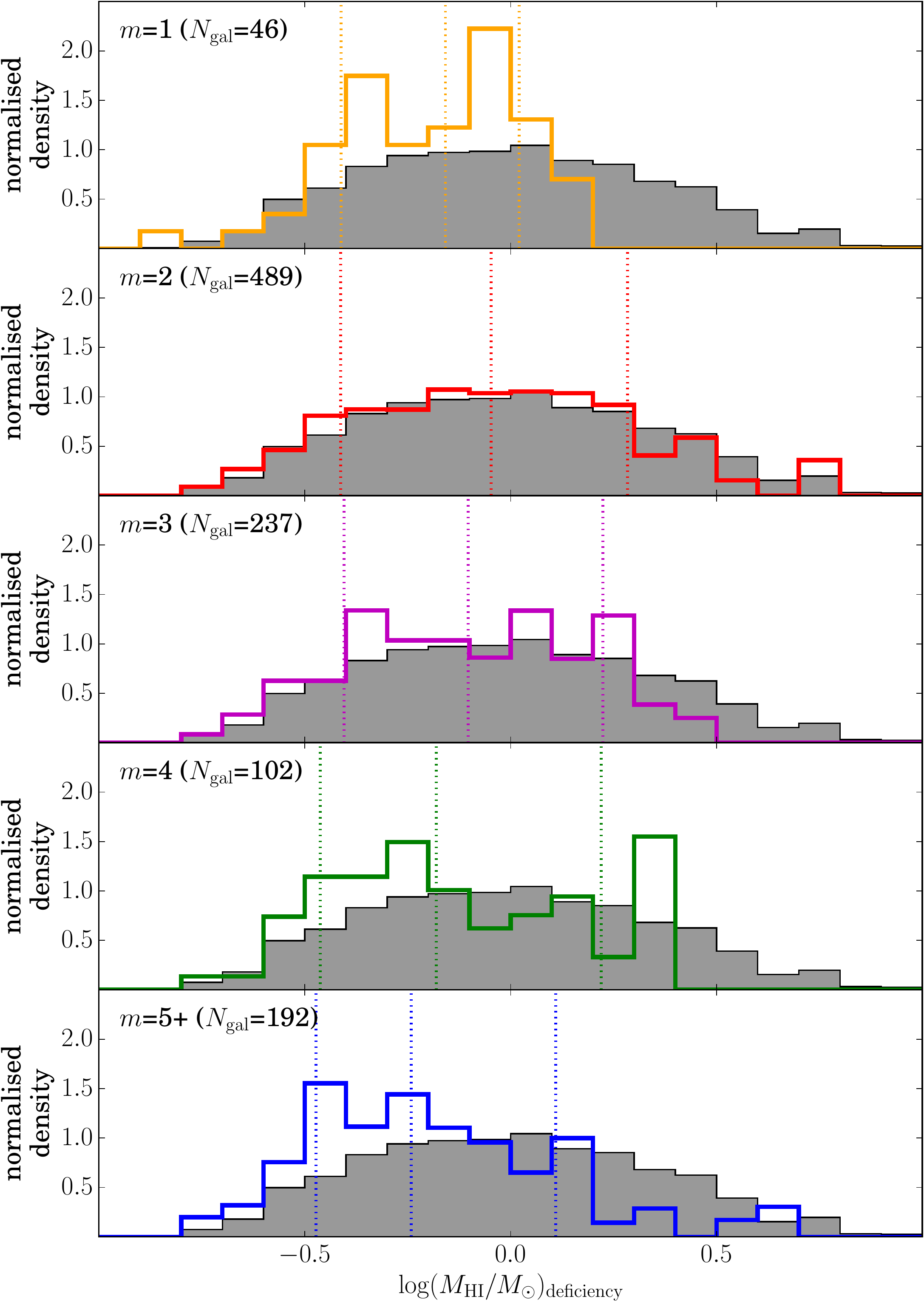}
    \caption{H\textsc{i} deficiency, calculated using equations \ref{eq:gas_expected} and \ref{eq:gas_deficiency}, for each of the \textit{arm number subsamples}. The underlying grey histograms show the distributions for all galaxies with detected H\textsc{i}, irrespective of morphology and the solid lines show the same distributions for all galaxies split by spiral arm number. Each H\textsc{i} detection is weighted by $1/V_\mathrm{max}$, and the vertical lines show the positions of the median, 16th and 84th percentiles for each spiral arm number. The dotted vertical lines show the median, 16th and 84th percentiles of each of the distributions.}
    \label{fig:gas_residual}
\end{figure}

To compare whether our spiral galaxy samples are deficient in gas for their stellar mass, we compare the measured gas fractions for the \textit{stellar mass-limited} sample of spirals, and the distributions are plotted in Fig.~\ref{fig:gas_residual}. Only galaxies with ALFALFA detections are included, giving 1,066 spiral galaxies in total. To ensure that a full range of gas masses is probed, we include all $\alpha$70 detections and apply a $V_\mathrm{max}$ weighting to the H\textsc{i} detections that fall below the H\textsc{i} complete mass of $\log(M_\mathrm{H\textsc{i}})$=9.89. The resulting H\textsc{i} deficiency distributions are plotted in Fig.~\ref{fig:gas_residual}. Here, we see that $m=2$ galaxies are more deficient in gas than many-armed galaxies. The median H\textsc{i} deficiencies are $-0.15 \pm 0.08$, $-0.05 \pm 0.02$, $-0.10 \pm 0.03$, $-0.18 \pm 0.05$ and $-0.24 \pm 0.04$ for $m=1$, $2$, $3$, $4$ and $5+$ respectively. We see a trend that many-armed spiral galaxy samples are more H\textsc{i} rich than $m=2$ galaxy samples. Although we cannot rule out the null hypothesis that the $m=3$ sample is from the same parent distribution, as the KS $p$-value is 0.31, it is unlikely that this is the case for the  $m=4$ and $m=5+$ \textit{arm number subsamples} with respect to the $m=2$ sample, where the corresponding $p$-values are $\sim 10^{-2}$ and $\sim 10^{-4}$.

\subsection{The role of bars}
\label{sec:bars}

\subsubsection{Bar fractions with arm number}
\label{sec:relative_bar_fractions}

One of the key features via which a grand design spiral pattern may emerge is via a  bar instability \citep{Kormendy_79}. The exact nature of the dependence of grand design spiral structure on the presence of a bar in a galaxy disc is not fully understood, since many-armed spiral galaxies can still exist in the presence of a bar, and not all grand design spiral galaxies host strong bars. Nonetheless, bars are more common in grand design spiral galaxies \citep{EE_82,EE_87}. Bars can affect the gas and star formation properties of their host galaxies \citep{Athanassoula_92,Oh_12,Masters_12}.  In previous sections, we have removed any galaxies with strong bars from the sample by only selecting galaxies with $P_\mathrm{bar} \leq 0.5$. To assess the impact that the presence of bars have on spiral galaxies and understand whether the differences in the galaxy populations are driven by the presence of bars with spiral structure, the properties of barred and unbarred galaxies are now compared. We include all spirals in this analysis, with no cut on $p_\mathrm{bar}$ for this section.

GZ2 has collected visual classifications of spiral galaxies, which have been used to define galaxies with and without bars. We use the same prescription of \citet{Masters_11} by selecting galaxies with $p_\mathrm{bar} > 0.5$ as barred. The fraction of galaxies with bars is significantly higher for the $m=2$ sample, with $50 \pm 1$ per cent of galaxies having bars, compared to $16$--$25$ per cent for each of the many-armed samples. This confirms the results of previous studies (e.g., \citealt{EE_82,EE_87}) for our sample: two-armed grand design spiral galaxies are more likely to host a bar than many-armed spiral galaxies.

\subsubsection{The effects of bars on star formation properties}
\label{sec:sf_w_bars}

In the previous sections, the total SFRs of spiral galaxies have been shown to differ little with respect to spiral arm number. However, more of the star formation is obscured in the $m=2$ population. The total gas fractions among all of the galaxies with a detection in $\alpha$70 also show that there is also a weak trend for many-armed spiral galaxies to be more H\textsc{i} rich. \citet{Masters_12} showed that the ALFALFA-measured H\textsc{i} fractions are much lower in barred galaxies, which may explain why fewer two-armed spiral galaxies have an H\textsc{i} detection in our sample. 
\begin{figure}
	
    \includegraphics[width=0.45\textwidth]{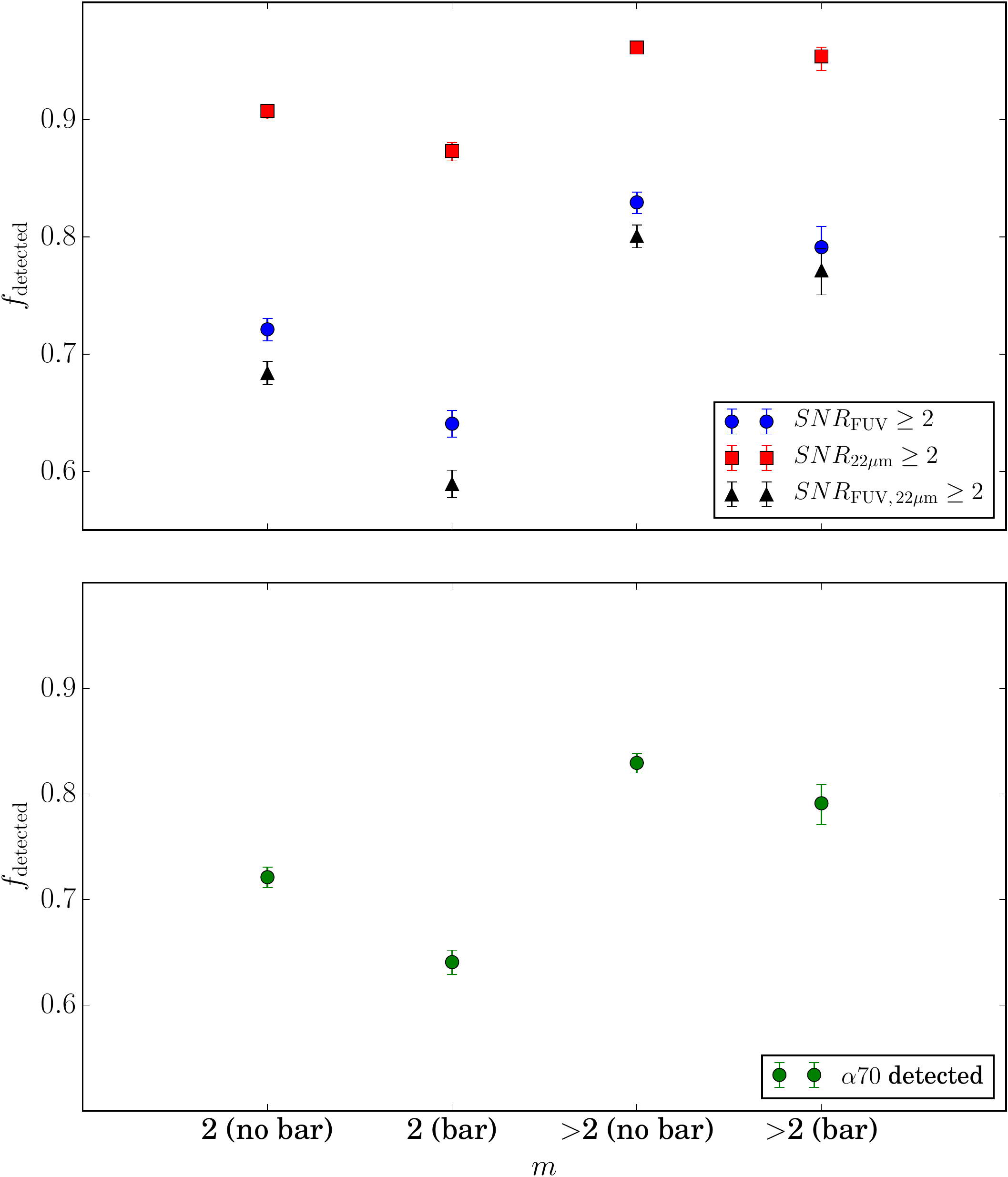}
    \caption{Top panel: fraction of galaxies with GALEX FUV and WISE $22\,\mathrm{\mu m}$ detections for each of the barred and unbarred spiral samples. Bottom panel: fraction of the galaxies in the $\alpha$70 survey region with a reliable  detection. The errorbars show the $\pm 1 \sigma$ errors, calculated according using the method of \citet{Cameron_11}.}
    \label{fig:b_completeness}
\end{figure}

Each of our many-armed samples comprise of fewer galaxies than the two-armed galaxy sample, and only a small number of those have bars. In order to compare the properties of barred and unbarred samples of galaxies with different arm numbers with good number statistics, we elect to group our 3, 4, and 5+ armed spiral galaxies. We deem this to be reasonable, since any trends seen in each of the many-armed spiral galaxy samples have been shown to be similar when compared to the $m=2$ sample. In this analysis, we include all galaxies classified as spiral, now including galaxies with $p_\mathrm{bar}>0.5$, which were removed for the earlier results. We first compare the completeness of the GALEX FUV, WISE $22\,\mathrm{\mu m}$ and $\alpha$70 in Fig.~\ref{fig:b_completeness}. Here we see that the overall completeness of each of these measures decreases for strongly barred galaxies, yet the detection fractions are still consistently higher for many-armed spirals. 

\begin{figure*}
	
    \includegraphics[width=0.975\textwidth]{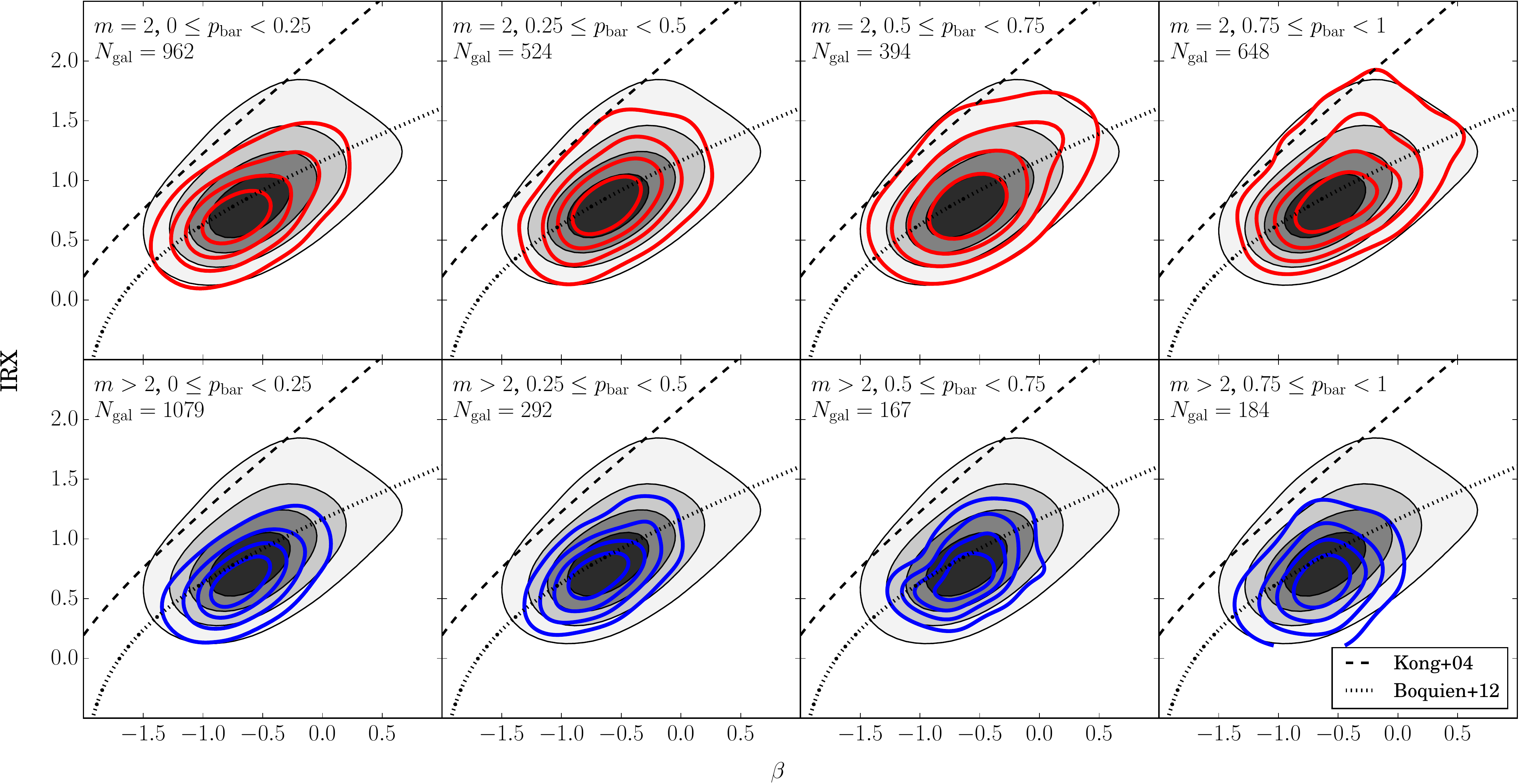}
    \caption{IRX-$\beta$ for $m=2$ and $m>2$ many-armed spiral galaxies with and without bars. The filled grey contour shows the distribution for all galaxies from the \textit{stellar mass-limited sample}, including galaxies of all morphologies. Four bins of GZ2 $p_\mathrm{bar}$ run from left to right for $m=2$ galaxies (top row) and $m>2$ galaxies (bottom row), and their respective points are indicated by the solid contours. The contours indicate the regions enclosing 20, 40, 60 and 80\% of the points for each sample. The measured IRX-$\beta$ relations for normal star-forming galaxies \citep{Boquien_12} and starburst galaxies \citep{Kong_04} are shown for reference.}
    \label{fig:irx_beta_w_bars}
\end{figure*}

In order to check whether the presence of bars in our galaxy \textit{arm number samples} affect the IRX-$\beta$ trends in Sec.~\ref{sec:irx_beta}, we subdivide the sample of spiral galaxies into four bins of bar strength, defined using the GZ2 $p_\mathrm{bar}$ statistic. The resulting IRX-$\beta$ relations for the $m=2$ and the $m>2$ samples are shown in Fig.~\ref{fig:irx_beta_w_bars}. The median offsets from the \citet{Boquien_12} relation are $-0.06 \pm 0.01$, $-0.02 \pm 0.02$, $0.00 \pm 0.02$ and $0.00 \pm 0.02$ for each of the bar strength bins for two-armed spirals. The corresponding offsets for the many-armed spirals are $-0.15 \pm 0.01$,
$-0.10 \pm 0.02$, $-0.11 \pm 0.02$ and $-0.17 \pm 0.03$. These results therefore show that with or without the presence of a bar, many-armed spirals are more UV luminous than expected for normal star-forming galaxies than two-armed spiral galaxies. 

It is evident from the top row of Fig.~\ref{fig:irx_beta_w_bars} that bars affect both IRX and $\beta$ in two-armed spirals, without causing any significant deviation from the IRX-$\beta$ relation for normal star-forming galaxies.

\section{Discussion}
\label{sec:discussion} 

\subsection{Total star formation rates}
\label{sec:discussion_sfr}

In Sec.~\ref{sec:sfms}, we compare the total SFRs of spiral galaxies with different arm numbers, finding only marginal differences between the samples. We note that galaxies with different spiral arm numbers occupy similar ranges of stellar mass (see Fig.~\ref{fig:mass_plot}) and find no enhancement in the measured SFR of the two-armed spirals relative to the many-armed spirals. Rather, galaxies with more spiral arms have slightly higher detection fractions (and hence less likelihood of very low SFR) in both the UV and MIR. For galaxies with secure measurements, 3- and 4-armed galaxies have marginally higher average sSFRs than those in the two-armed sample.

Many-armed spiral patterns occur readily in  simulations of undisturbed discs, and tend to be transient in nature \citep{Bottema_03,Baba_09,Grand_12b,Baba_13,Donghia_13,Roca-Fabrega_13}. In contrast, stable two-armed spiral patterns usually require some form of perturbation \citep{Sellwood_11}, in the form of a tidal interaction, bar instability or density wave. An enhancement in the current SFR would be expected if density waves were responsible for the triggering of star-formation \citep{Roberts_69}, bars were triggering star formation in the galaxy centre \citep{Athanassoula_92} or if there were ongoing tidal interactions \citep{Barton_00,Ellison_08} in two-armed galaxies. Such mechanisms should not be evident in many-armed galaxies. Our results show no strong evidence for any SFR enhancement in two-armed spiral galaxies. These results therefore support those of \citet{EE_86}, \citet{Stark_87} and \citet{Willett_15}, where it was found that different forms of spiral structure do not lead to a deviation from the total SFR relations of local galaxies. They are also consistent with \citet{Foyle_11} and \citet{Choi_15}, where there was no conclusive evidence for the triggering of star formation by grand design spiral arms themselves. Instead, our results favour a scenario where spiral arms simply reflect the arrangement of star-forming material in galaxies, without being directly responsible for the triggering of star formation \citep{Vogel_88,Elmegreen_02,Moore_12}.


\subsection{Obscured vs. unobscured star formation}
\label{sec:discuss_uv_vs_ir}

Spiral arms are regions of both increased star formation and dust obscuration \citep{Helou_04,Calzetti_05,Grosbol_12}.  In Sec.~\ref{sec:uv_vs_mir} we find that, although overall SFRs are seemingly unaffected by spiral arm number, the fraction of the young stars that are obscured by dust differs significantly.

At a given star formation rate, two-armed spirals display more MIR dust emission, indicating that a greater proportion ($\sim 10$ per cent) of their young stellar population resides in heavily obscured regions.  This is likely due to different relative distributions of star-formation and dust in galaxies with different numbers of spiral arms.  

If we consider the IRX-$\beta$ diagram, we see that (unbarred) galaxies with all numbers of arms have similar $\beta$ distributions, indicating that the amount of extinction affecting the \emph{observed} UV light does not vary substantially with spiral structure.  On the other hand, IRX varies substantially, indicating more extinction for two-armed spirals.  In order to avoid modifying $\beta$, this additional extinction must be in the form of heavily obscured regions, from which almost no UV escapes.  Therefore, it is the relative distribution of young stars and regions of high extinction which varies with arm number.

However, regions of very high dust extinction are the same places in which stars form, and so their distributions are closely related.  A number of possibilities seem to be admitted by our results.  Consider a single star-forming molecular cloud within a galaxy spiral arm.  The fraction of young stars which are heavily obscured could depend on the covering fraction of surrounding, but otherwise unrelated, molecular clouds.  Alternatively, it may depend on the degree to which the young stars have escaped their own birth cloud.  In the first case, the obscured fraction is determined primarily by geometry: by the relative spatial distribution (and perhaps sizes) of star-forming regions. In the second scenario, the obscured fraction is related to the time since the region commenced star-formation, and perhaps other physical properties of the molecular cloud.

Grand design spiral arms are typically better defined and higher-contrast than many-arm structures \citep{Elmegreen_11b}.
We have also seen that, despite having fewer arms, they have similar total star-formation rates. These observations imply that two-arm spirals have more, or larger, star formation regions in a given volume of spiral arm. 
This could result from mechanisms associated with grand design spiral structure, such as a strong density wave, that act to gather more gas and dust into spiral arm regions.  For example, simulations suggest that the presence of density waves leads to the formation of more massive molecular clouds \citep{Dobbs_11,Dobbs_12}.
Since the molecular clouds present a larger local cross section, a greater fraction of young stars are obscured by dust. The ratio of MIR to UV emission (i.e., IRX) is therefore higher.

Alternatively, the obscured fraction may be related to the recent star-formation history. If this is more bursty in nature, then more of the resulting luminosity is emitted in the MIR than in the UV: the IRX-$\beta$ relation is displaced upwards \citep{Meurer_99,Kong_04}. If star-formation in grand design spirals is driven by a triggering mechanism -- such as a tidal interaction with a companion galaxy \citep{Sundelius_87,Dobbs_10} or a density wave \citep{Seigar_02,Kendall_15} -- while many-armed spirals are not subject to the same mechanisms, then one would expect their recent star-formation histories to differ. \citet{Boquien_12} show that the scatter in IRX-$\beta$ for star-forming galaxies is attributable to the intrinsic UV slope ${\beta}_0$, which is sensitive to the recent star formation history. \citealt{Kong_04} suggests that for a period of $\sim 1$\,Gyr \emph{following} a starburst, galaxies will occupy a lower position in the IRX-$\beta$ plane: as the new stars escape their molecular birth clouds the galaxy becomes UV bright, and the declining MIR emission is not replaced by further star formation. As many-armed spirals lie lower in the IRX-$\beta$ plane, it is possible they are associated with a (mild) post-starburst state. In \citet{Hart_16}, we showed that a recent, rapid decline in SFR was required to produce the observed optical colours for many-armed spirals, which would be consistent with these observations. 

A further possibility is that the dispersal time of molecular clouds varies with spiral arm number. Although both UV and MIR emission are associated with recent star formation (e.g., \citealt{Hao_11,Kennicutt_12}), they actually probe different timescales in the evolution of star-forming regions. In nearby galaxies, UV emission is displaced from the H$\alpha$ emission that traces the most recent star formation, whereas MIR emission arises from regions much closer to the brightest  H$\alpha$ knots \citep{Helou_04,Calzetti_05,Crocker_15}. To observe UV emission, the parental molecular cloud must be dispersed, a process that takes $\sim 7$\,Myr in local spirals \citep{Grosbol_12}. However, more massive molecular clouds may take longer to disperse. The dispersion of molecular clouds may thus be a more rapid process in many-armed galaxies, with weaker spiral structure, meaning that the UV emitting population emerges more quickly.

The radial geometry of the star formation in spiral galaxies may also be affected by the presence of differing forms of spiral structure. Since the level of dust obscuration is greater towards the centre of galaxies (e.g., \citealt{Boissier_04,Roussel_05,Boissier_07}) these results could imply that star-formation occurs more centrally in two-armed spiral galaxies, which is proposed to be the case for barred spirals (e.g., \citealt{Athanassoula_92,Oh_12}). Given that bars are also associated with two-armed spiral structure \citep{Kormendy_79}, this may have a strong influence on the observed properties of galaxies. However, in Sec.~\ref{sec:sf_w_bars} we investigated what effect the presence of a strong bar has on the IRX-$\beta$ diagram, and found that the presence of bars in our galaxies led to no differences in the offset from the IRX-$\beta$ relation. This  suggests that the spiral structure itself is responsible for the observed offsets from the normal IRX-$\beta$ relation, rather than bars and the radial geometry of star formation.

Discerning between the possibilities outlined above will require a more careful consideration of the distribution, properties and evolution of molecular clouds within individual galaxies.  However, with further synthesis of empirical results and sophisticated modelling, further progress in understanding how spiral arms relate to star-formation seems promising.

\subsection{HI fractions}
\label{sec:discussion_gas}

Gas plays a critical role in sustaining spiral structure in discs.  Gas in discs contributes to the growth of spiral perturbations via swing amplification in both grand design and many-armed spirals \citep{Jog_92,Jog_93}.  The accretion of cool gas onto galaxy discs can also help sustain many-armed patterns.  The role that the amount of gas plays in galaxies with different spiral arm numbers was considered in Sec.~\ref{sec:gas}.  Given that gas could potentially amplify or sustain both two-armed \citep{Ghosh_15,Ghosh_16} and many-armed spiral structure \citep{Jog_93}, then it is expected that all of our galaxy samples should contain significant quantities of H\textsc{i}. We found that many-armed spiral galaxies are the most gas rich, whereas two-armed spirals are the most gas deficient. Given that the total SFRs are similar for all of our samples, this implies that H\textsc{i} is converted to stars more efficiently in two-armed spirals than in many-armed spirals. As swing amplification acts to amplify all types of spiral structure, it is unclear why different spiral arm patterns would be more or less gas rich. It is perhaps the case that a higher gas fraction is required to sustain a many-armed spiral pattern in galaxy discs.

Another factor to be considered is the presence of bars in our spiral galaxies.  In our analysis, we remove all strongly barred galaxies, yet still see a trend for two-armed galaxies, which are usually associated with bar instabilities, to be the most gas poor \citep{Davoust_04,Masters_12}. Therefore, variation in the gas fraction must also relate directly to differences in spiral arm structure or to the presence of weak bars.

\section{Conclusions}
\label{sec:conclusions}

In this paper, the star formation properties of spiral galaxies were investigated with respect to spiral arm number. A sample of galaxies was selected from SDSS, using the visual classification statistics of Galaxy Zoo 2 (GZ2). Using photometry from GALEX, SDSS and WISE, total star formation rates were compared by combining measures of unobscured star-formation from UV emission and obscured star-formation from the MIR. Many-armed spiral galaxies are less likely to have very low sSFRs and thus be undetected in the UV or MIR. However, for galaxies with reliable UV and MIR detections, sSFR has no significant dependence on spiral arm number. Despite this, we find that spirals with different numbers of arms do have different levels of dust obscuration, with many-armed spirals having more UV emission from young stars unobscured by dust. This is most evident when comparing the IRX-$\beta$ relation, where IRX measures the relative fraction of MIR to UV emission, and $\beta$ is the UV slope. Many-armed spirals have significantly lower IRX for a given $\beta$. We have discussed several possibilities that could give rise to our findings, individually or in combination:
\begin{enumerate}
\item The differences could be due to the relative distribution of star forming regions in galaxies with different spiral structure. In grand design spirals, the molecular clouds may be more concentrated in the dense arm regions. The consequent increase in dust obscuration may lead to a reduction in UV emission compared to that in the MIR.

\item More generally, molecular cloud properties, e.g. mass and size distributions, may differ in discs hosting two or more spiral arms.  In this case, in addition to geometrical effects, molecular clouds may take longer to disperse in two-armed galaxies. The UV emitting population would thus emerge over a longer timescale, leading to an enhanced IRX.

\item Our results also appear consistent with many-armed spirals having recently experienced a rapid decline in star-formation rate.
\end{enumerate}

Two-armed spirals are more gas deficient than many-armed spirals, meaning they are more efficient at converting  to stars. Two-armed spirals are also more likely to host bars, with $\sim 50$ per cent having strong bars compared to only $\sim 20 $ per cent of many-armed galaxies. However, we show that bars only serve to move galaxies along the normal IRX-$\beta$ relation: strongly barred galaxies have higher levels of MIR emission as well as steeper UV slopes.  Spiral arm number, on the other hand, has a significant effect on how far spirals are offset perpendicular to the normal IRX-$\beta$ relation.

\section*{Acknowledgements}

The data in this paper are the result of the efforts of the Galaxy Zoo 2 volunteers, without whom none of this work would be possible. Their efforts are individually acknowledged at \url{http://authors.galaxyzoo.org}.

The development of Galaxy Zoo was supported in part by the Alfred P. Sloan foundation and the Leverhulme Trust.

RH acknowledges a studentship from the Science and Technology Funding Council, and support from a Royal Astronomical Society grant. 

Plotting methods made use of \texttt{scikit-learn} \citep{scikit-learn} and \texttt{astroML} \citep{astroML}. This publication also made extensive use of the \texttt{scipy} Python module \citep{scipy},  and \texttt{TOPCAT} \citep{topcat} and \texttt{Astropy}, a community-developed core Python package for Astronomy \citep{astropy}. 

Funding for SDSS-III has been provided by the Alfred P. Sloan Foundation, the Participating Institutions, the National Science Foundation, and the U.S. Department of Energy Office of Science. The SDSS-III web site is http://www.sdss3.org/.

SDSS-III is managed by the Astrophysical Research Consortium for the Participating Institutions of the SDSS-III Collaboration including the University of Arizona, the Brazilian Participation Group, Brookhaven National Laboratory, Carnegie Mellon University, University of Florida, the French Participation Group, the German Participation Group, Harvard University, the Instituto de Astrofisica de Canarias, the Michigan State/Notre Dame/JINA Participation Group, Johns Hopkins University, Lawrence Berkeley National Laboratory, Max Planck Institute for Astrophysics, Max Planck Institute for Extraterrestrial Physics, New Mexico State University, New York University, Ohio State University, Pennsylvania State University, University of Portsmouth, Princeton University, the Spanish Participation Group, University of Tokyo, University of Utah, Vanderbilt University, University of Virginia, University of Washington, and Yale University.

We thank the many members of the ALFALFA team who have contributed to the acquisition and processing of the ALFALFA data set.




\bibliographystyle{mnras}
\bibliography{bibliography} 




\bsp	
\label{lastpage}
\end{document}